\documentclass{aastex62}

\graphicspath{{./}{figures/}}

\received{\today}
\revised{\today}
\accepted{\today}
\submitjournal{ApJ}

\shorttitle{ALMA monitoring of mm line variation in IRC+10216}
\shortauthors{He et al.}

\begin{document}

\title{ALMA monitoring of mm line variation in IRC\,+10216: I. Overview of millimeter variability.}

\correspondingauthor{J. H. He et al}
\email{jinhuahe@ynao.ac.cn}

\author[0000-0002-3938-4393]{J. H. He}
\affil{Yunnan Observatories, Chinese Academy of Sciences, 396 Yangfangwang, Guandu District, Kunming, 650216, P. R. China}
\affiliation{Chinese Academy of Sciences South America Center for Astronomy, China-Chile Joint Center for Astronomy, Camino El Observatorio \#1515, Las Condes, Santiago, Chile}
\affiliation{Departamento de Astronom{\'i}a, Universidad de Chile, Casilla 36-D, Santiago, Chile}

\author{T. Kami{\'n}ski}
\affiliation{Center for Astrophysics, Harvard \& Smithsonian, 60 Garden Street, Cambridge, MA, USA}
\author{R.~E. {Mennickent}}
\affiliation{Universidad de Concepci{\'o}n, Departamento de Astronom{\'i}a, Casilla 160-C, Concepci{\'o}n, Chile}
\author{V. I. Shenavrin}
\affiliation{Sternberg Astronomical Institute, Moscow M.V. Lomonosov State University, Universitetskij pr., 13, Moscow, 119992, Russia}
\author{Diego {Mardones}}
\affiliation{Departamento de Astronom{\'i}a, Universidad de Chile, Casilla 36-D, Santiago, Chile}
\author{Wei Wang}
\affiliation{Chinese Academy of Sciences South America Center for Astronomy, China-Chile Joint Center for Astronomy, Camino El Observatorio \#1515, Las Condes, Santiago, Chile}
\author{Baitian Tang}
\affiliation{School of Physics and Astronomy, Sun Yat-sen University, Zhuhai 519082, Peopleʼs Republic of China}
\author{M.~R. {Schmidt}}
\affiliation{N. Copernicus Astronomical Center, Rabia{\'n}ska 8, 87-100 Toru{\'n}, Poland}
\author{R. {Szczerba}}
\affiliation{N. Copernicus Astronomical Center, Rabia{\'n}ska 8, 87-100 Toru{\'n}, Poland}
\author{Jixing. {Ge}}
\affiliation{Departamento de Astronom{\'i}a, Universidad de Chile, Casilla 36-D, Santiago, Chile}

\begin{abstract}

Temporal variation of millimeter lines is a new direction of research for evolved stars. It has the potential to probe the dynamical wind launching processes from time dimension. We report here the first ALMA (Atacama Large Millimeter Array) results that cover 817 days of an on-going monitoring of 1.1\,mm lines in the archetypal carbon star IRC\,+10216. The monitoring is done with the compact 7-m array (ACA) and in infrared with a 1.25\,m telescope in Crimea. A high sensitivity of the cumulative spectra covering a total of $\sim7.2$\,GHz between 250 -- 270\,GHz range has allowed us to detect about 148 known transitions of 20 molecules, together with more of their isotopologues, and 81 unidentified lines. An overview of the variabilities of all detected line features are presented in spectral plots. Although a handful of lines are found to be very possibly stable in time, most other lines are varying either roughly in phase or in anti-correlation with the near-infrared light. 
Several lines have their variations in the ALMA data coincident with existing single dish monitoring results, while several others do not, which requires an yet unknown mechanism in the circumstellar envelop to explain. 

\end{abstract}



\section{Introduction} \label{sec:intro}

Asymptotic Giant Branch (AGB) stars are fascinating stars that have significant contributions, through their intense stellar winds, to the cycling of heavy elements, dust grains and a wide array of molecules in the interstellar medium. Although the AGB stellar winds have long been thought to be driven by radiation pressure acting on dust grains \citep{Kwok1975}, direct observations of the tiny dust-forming wind-launching regions around the AGB stars are still rare, except for a few interferometric observations \citep[sometimes combined with polarization; see e.g.,][]{Sacu2013,Deci2015,Ohna2016,Ohna2017,Ohna2019}. Therefore, the development of diverse observational probes to the AGB wind acceleration zones is sorely needed.

Molecular line variability is one such powerful probe to the wind acceleration zones around AGB stars, because the variation of many molecular lines originates in that region. Strong maser lines have been the major proxies for variability studies of (mainly oxygen-rich) AGB stars at radio wavelengths for a long time. But, of well-studied lines, only the SiO masers are pumped close enough to the central AGB star to allow probing the wind acceleration zone. Very long baseline intereferometry (VLBI) mapping has uncovered clumpy ring like patterns of SiO masers around evolved stars \citep[e.g.,][]{Alco1986,McIn1989,Doel1998}. However, even the multi-epoch VLBI mapping of the SiO masers \citep[e.g.,][]{Imai2010} can tell us little about the physical conditions of the dust forming wind launching region, because the maser spots only sample very small volumes. Therefore, high spatial resolution temporal monitoring of non-maser lines is indispensable to probing the AGB wind launching processes. 

The variability of millimeter non-maser molecular lines of AGB stars were only investigated since recently by two pioneering single dish studies of \citet{Cern2014} and \citet{He2017}, both targeting the same source, IRC\,+10216. The former made use of the repeated observations by Herschel Space Observatory that has a better flux calibration accuracy than ground based single dishes. The latter used a ground based single dish telescope but relied on a relative flux-calibration approach using stable and varying lines observed in the same spectral band pass (their so called "in-band calibration" method). A new single dish monitoring program using the IRAM\,30m telescope had been initiated by the first team right after the publication of their Herschel paper. Their first results have been reported in \citet{Fonf2018} and \citet{Pard2018}. Our team also has upgraded our monitoring program from single dish to the ALMA interferometer since 2016 to benefit from a higher spatial resolution. The first results are reported in this paper.

The thus-far discovered variations in molecular transitions are complex. Firstly, line variations appear in both maser and non-maser components of the lines and the line strengths can change by up to $50\%$. \citet{He2017} advocated the maser nature of SiS,14-13 and HCN,$v_2$=1$^f$,3-2 lines based on the analysis of the variation amplitudes and velocities, while \citet{Fonf2018} revealed the maser nature of SiS,14-13 and 15-14 lines with the aid of high spatial resolution maps from CARMA and ALMA. 
Second, while many lines vary roughly in phase with the near infrared (NIR) light, some other lines show large phase shifts against the NIR light and the shifts can amount to about half NIR period. The former group includes multiple transitions of SiS \citep{He2017,Fonf2018}, C$_4$H lines in the $v_7$ vibrational states \citep{He2017}, CN and most other lines presented by \citet{Pard2018}; the latter group involves CCH and C$_4$H lines in the ground vibrational state \citep{Pard2018}, $^{30}$SiO,6-5, HCN,$v_2$=1$^f$,3-2 and a blended line feature of Na$^{37}$Cl, CH$_2$NH, and HC$_3$N \citep{He2017}. 
Third, the variation period, phase and relative amplitude can be velocity dependent within a single line profile \citep{He2017} and non-Gaussianity can appear in the light curves \citep{He2017,Fonf2018}. 
Finally, most of the varying line emission can be attributed to the central hot and dense wind-acceleration zone within $r\lesssim 11$ stellar radii, while only the varying CCH and C$_4$H lines in the ground state could be dominated by the emission from the outer region (around a radius of $15''$) because of their large phase delay and known ring like emission structure. 

In this work, we present the average spectra and line identification, and give a qualitative presentation of the line variability of all lines identified in the ALMA monitoring program. We also uncover the evidence of radial dependence of the mm line variations. We first describe our ALMA and NIR observations in Sect.~\ref{sec:observations}, and then data reduction and the current difficulties in constraining flux uncertainties in Sect.~\ref{sec:data reduction}. The results are discussed in Sect.~\ref{sec:Results} and summarized in Sect.~\ref{sec: summary}.

\section{Observations} \label{sec:observations}

Our observations consist of two parts: the ALMA monitoring and a NIR multi-band photometric monitoring. The 7\,m antennas of the Atacama Compact Array (ACA) is used for the monitoring because its interferometric baselines do not change with time drastically. The observations were always performed at an elevation angle of about $50^{\circ}$ to minimize the interferometric effects of changing projected baselines that would be very harmful to sensitive monitoring. The monitoring started in 2016 and have been executed for three proposal cycles (with ALMA project codes: 2016.1.00794.S, 2016.2.00033.S, 2017.1.01689.S and 2018.1.00047.S); the last cycle is still ongoing in 2019. We report the initial results from the first 21 epochs (see the details in Table~\ref{tab: epochs}) that cover 817 days ($\sim2.2$\,yrs), which is $\sim 1.3$ times of the pulsation period of 630\,day ($\sim1.7$\,yrs) of the target IRC\,+10216 \citep{Ment12}. The typical on-source integration time was $\sim4$\,min in Cycle 4, but increased to $\sim5$\,min in Cycle 5 and $\sim10$\,min in Cycle 6, which results in a typical root mean square (RMS) noise of $\sim 25$\,mJy/beam ($\sim 10$ to $\sim 75$\,mJy/beam) at the effective spectral resolution of $\sim 1$\,MHz ($\sim 1.2$\,km\,s$^{-1}$) in the spectral baselines. The change of weather conditions has significant contribution to the difference in RMS levels. The spectral resolution is sufficient to get the typical line width of $\sim 29$\,km\,s$^{-1}$ well resolved. There are four spectral windows, with each covering about 1.8\,GHz frequency width. The observed frequency ranges varied slightly among epochs due to Doppler correction; the overlapped ranges in lab frame (with the target velocity of -26.5\,km\,s$^{-1}$ corrected) are about 251293-253105, 253594-255406, 265791-267603 and 268192-270018\,MHz. In addition to the two varying lines of HCN and SiS that were monitored in our single dish work \citep{He2017}, the ACA monitoring covers many more strong transitions from various vibrational states. The gain calibrators were almost always J0854+2006 (for both amplitude and phase, sometimes also for band pass) and J1058+0133 (mainly for amplitude and band pass). The latter also serves as the flux calibrator in most cases. Only in rare cases, other calibrators (J1041+0610, J0909+0121, J1037+2934, J0750+1231 and J0830+2410) were also used. A single 7\,m antenna has a primary beam of about $42''$, while the synthesized beam size and maximum recoverable scale of ACA is $\sim4''\times 8''$ and $28''$, respectively. Some lines such as those of C$_4$H, SiCC, c-C$_3$H$_2$ and HC$_3$N have extended emission that could have been spatially filtered out by ACA. The array configuration is relatively stable, with 9-12 antennas available during the observations.
The monitoring cadence was about once every two months in the first year and was enhanced to once a month since the second year so that we have about 10-20 epochs per stellar pulsation period.
\begin{deluxetable}{llll}[b!]
\tablecaption{ALMA observations.\label{tab: epochs}}
\tablecolumns{4}
\tablenum{1}
\tablewidth{0pt}
\tablehead{
\colhead{Epoch} &
\colhead{OUS\tablenotemark{a}} &
\colhead{Obs.Date.Time\tablenotemark{b}} &
\colhead{JD\tablenotemark{c}}
}
\startdata
01. & A001/X891/X33   & 2016-10-21 11:56:06     & 2457682.994534  \\
02. & A001/X891/X37   & 2017-01-21 05:12:48     & 2457774.714472  \\
03. & A001/X891/X3b   & 2017-03-27 02:40:51     & 2457839.608909  \\
04. & A001/X891/X3f   & 2017-04-13 00:52:22     & 2457856.533581  \\
05. & A001/X891/X43   & 2017-06-22 18:54:30     & 2457927.285092  \\
06. & A001/X1262/X51  & 2017-07-04 19:28:22     & 2457939.305184  \\
07. & A001/X891/X47   & 2017-08-07 17:44:36     & 2457973.236523  \\
08. & A001/X1296/X846 & 2017-10-04 13:28:00     & 2458031.057982  \\
09. & A001/X1296/X849 & 2017-11-04 11:43:49     & 2458061.985619  \\
10. & A001/X1296/X84c & 2017-12-04 10:59:44     & 2458091.955005  \\
11. & A001/X1296/X84f & 2018-01-04 07:30:28     & 2458122.809720  \\
12. & A001/X1296/X852 & 2018-03-20 03:00:18	    & 2458197.622098  \\
13. & A001/X1296/X855 & 2018-04-18 00:46:58	    & 2458226.529522  \\
14. & A001/X1296/X858 & 2018-05-18 00:46:58	    & 2458254.466304  \\
15. & A001/X1296/X85b & 2018-07-10 18:46:53	    & 2458310.279422  \\
16. & A001/X1296/X85e & 2018-08-12 17:51:24	    & 2458343.240888  \\
17. & A001/X1296/X861 & 2018-09-09 15:34:08	    & 2458371.162682  \\
18. & A001/X133d/X5ae & 2018-10-16 13:55:02	    & 2458408.074014  \\
19. & A001/X133d/X5b1 & 2018-11-16 09:54:58	    & 2458438.907334  \\
20. & A001/X133d/X5b4 & 2018-12-19 08:30:25	    &           2458471.848615  \\
21. & A001/X133d/X5b7 & 2019-01-16 06:57:42	    &       2458499.784213  \\
\enddata
\tablenotetext{a}{OUS = ALMA Member Observing Unit Set that is observed with a single scheduling block.}
\tablenotetext{b}{The observation date and time is the end time of the observation shown in SnooPI in the ALMA webpage.}
\tablenotetext{c}{The JD is the mean JD of the whole scheduling block (SB) and is extracted from the calibrated measurement set using the {\em lstrange} function of the Python module {\em analysisUtils} which can be downloaded from the CASA web page \url{https://casaguides.nrao.edu/index.php/Analysis_Utilities}.}
\end{deluxetable}

The NIR photometric monitoring in J, H, K, L and M bands is performed with the Crimea 1.25\,m telescope of the Observatory of Sternberg Astronomical Institute, Lomonosov Moscow State University.  It started later than the ALMA monitoring program by about 1 year, with the first epoch on Dec. 12, 2017. A single photovoltaic InSb detector is used in a beam switching mode ($30''$ chops). The telescope system is stable, with a typical uncertainty of about 0.1\,mag or less in all the five bands. See more details of the telescope and observation in \citet{Shen11}.

\section{Data reduction} \label{sec:data reduction}

We use data calibrated by ALMA staff using default pipelines. Common Astronomy Software Applications (CASA 5.3.0-143) package\footnote{CASA is developed by an international consortium of scientists based at the National Radio Astronomical Observatory (NRAO), the European Southern Observatory (ESO), the National Astronomical Observatory of Japan (NAOJ), the Academia Sinica Institute of Astronomy and Astrophysics (ASIAA), the CSIRO division for Astronomy and Space Science (CASS), and the Netherlands Institute for Radio Astronomy (ASTRON) under the guidance of NRAO.} \citep{McMu2007} is used for imaging. A map of $60''\times 60''$ ($\sim1.4$ times of the primary beam) is produced for cleaning with {\em tclean}. The pixel size is set to $1.2''$, which over-samples the synthesized beam by a factor of $\sim5$. The {\em hogbom} deconvolver and {\em briggs} weighting (with parameter robust=0.5) are used. There can be different strategies to define cleaning boxes (e.g., primary beam masking, manual masking and auto masking). In this first report, we only adopt the primary beam masking at a level of {\em pbmask=0.2}, while the impact of the different masking strategies to the measured light curves will be tested in near future when we quantitatively analyze the light curves. Some molecular line features (e.g., the HCN and SiS lines) can be very strong. If we clean a data cube (e.g., all $\sim 1800$ frequency channels of a spectral window) as a whole, as in the standard procedure in CASA, these strong lines may prevent the convergence of {\em tclean} in other fainter channels because of the sharing of some convergence criteria such as flux threshold. To avoid this problem, we utilize the {\em tclean} to each individual channel to produce single channel clean maps and then concatenate the cleaned single channel maps back into an image cube for each spectral window using the CASA command {\em ia.imageconcat}. The flux threshold for the major cycles of {\em tclean} is set to $5\,\sigma$ level. However, the RMS of the image pixels ($\sigma$) initially measured from the whole image plane before the start of {\em tclean} can be significantly different for different frequency channels, depending on the level of line emission in the channels. This may harm the study of the variation of spectral line shapes in this project. In order to achieve as uniform cleaning as possible for all channels, we repeatedly measure $\sigma$ from the residual map after a standard {\em tclean} procedure has finished and use a loop over {\em clean} to achieve the convergence of $\sigma$ for each channel.  

The variability studies are very sensitive to  {\it tclean} procedures. Although the flux uncertainties of strong lines may be dominated by flux calibration uncertainty, which is several per cent in ALMA Band\,6 (we can adopt $8\%$), the relative variability among different lines, between the lines and continuum, among the frequency channels within a line, and between different sub-regions of the clean image is largely immune of the flux calibration uncertainty, because it only produces a systematic offset. Instead, other sources of uncertainty such as instrumental and atmospheric noises and inferometric defects (mostly side lobe effects) become important. Unfortunately, strict error transfer from UV plane to image plane is not yet included in CASA. The conventional method of measuring the flux uncertainty from an apparently emission-free region of the clean map also suffers from the neglect of the inevitable pixel correlations and non-uniformity of the uncertainties among pixels. A novel maximum likelihood approach of interferometric imaging that has the potential to resolve all these issues is already under developing by our team (in the step of prototyping) and will be presented soon in our future papers. Therefore, we mainly report qualitative results of our monitoring in this first report and leave the more quantitative analysis for future work using the new imaging approach.

It is possible to do self calibration to our data using the continuum. 
However, this procedure may introduce some degree of arbitrariness into the calibrated light curves, whose impact to our variability study is subject to scrutiny. We would like to leave this step for the future quantitative analysis steps. Therefore, the data presented in this paper has no any additional calibration aside from the standard processing in ALMA pipelines.

Concerning the NIR photometry, a detailed description of the data reduction can be found in \citet{Shen11}. In this work, we will only use the {\em K}-band light curve as an indicator of the central star pulsation to compare with the 1.1\,mm continuum and spectral line variations. {\em K}-band is close to the peak of the spectral energy distribution of the star and traces well the changes in bolometric luminosity of the star. 

\section{Results} \label{sec:Results}

\subsection{Average spectra and line identification} \label{subsec:ave_spec_line_identification}

\begin{figure}
    \centering
    \includegraphics[scale=0.5,trim={0 75 0 55},clip]{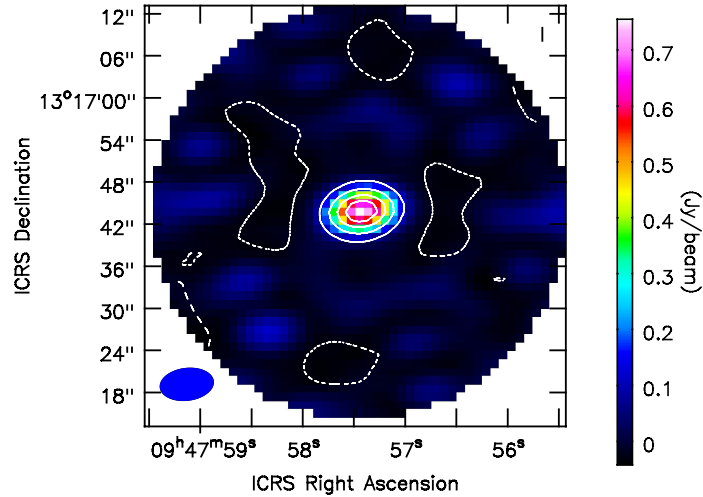}
    \caption{The integrated strength (moment 0 map) of SiS,14-13 line (contour) overlaid on 1.1\,mm continuum map (color scale; from only one of the four 1.8\,GHz spectral windows) for the first epoch of the ACA monitoring. The contour levels are -5, 20, 40, 60 and 80\% of the peak line strength of 2840\,Jy/beam$\times$km\,s$^{-1}$. The primary beam correction is not applied to this figure. The synthesized beam of about $4''\times 8''$ is indicated at the bottom left corner.}
    \label{fig:map}
\end{figure}

The 1.1\,mm continuum and the emission in most molecular lines in the ALMA data are limited to a point source toward the stellar position of IRC\,+10216 at ICRS\,09:47:57.4060 +13.16.43.560, except a handful of lines (e.g., some lines of C$_4$H and SiCC) that show a well known ring like morphology with a radius of around $13''$ \citep[e.g.][among others]{Guil1987,Gens1992,Taka1992,Daya1993,Guel1993,Luca1995,Cook15}. As an example, we show in Fig.~\ref{fig:map} the continuum map obtained in one spectral window of the first epoch and integrated line map of SiS,14-13 (contours). This line is a representative case of many others that show centrally peaked emission. Some other lines such as those of C$_4$H and SiCC also show a ring like emission component at a radius of about $13''$. In this paper, we only present spectra that are extracted from a single synthesized beam toward the central star position (averaged among the pixels in the central $2.4''\times2.4''$ square region to guarantee the inclusion of the strongest pixel for centrally peaked cases). The variation of the extended emission with time will be investigated in near future when the new imaging approach is available. Figure~\ref{fig: spec_ave} shows the spectra averaged over all 21 epochs (with the continuum included). Because each of the four 1.8\,GHz spectral windows is too wide for display, we divide each of them into two chunks of roughly equal widths. Each of the eight sub-plots consists of three panels that sequentially zoom in to the weaker line features from top to bottom. Lab frequency is used, which means the Doppler velocity of the star is zero in these plots.

The stacking of the repeated observations has achieved a deep baseline RMS of $\sim 10$\,mJy/beam per 1\,MHz channel, which enables the detection of new weak lines. The software Centre d'Analyse Scientifique de Spectres Instrumentaux et Synthétiques (CASSIS, version 5.0, with line list database cassis20161026.db)\footnote{The CASSIS web page is at \url{http://cassis.irap.omp.eu/}.} is used for line identification. In addition, the splatalogue\footnote{The splatalogue web is at \url{http://www.cv.nrao.edu/php/splat/}.}, Cologne Database for Molecular Spectroscopy \citep[CDMS;][]{Mull01,Mull05,Endr2016}\footnote{The CDMS webpage is at \url{https://cdms.astro.uni-koeln.de}.} and Jet Propulsion Laboratory line list \citep[JPL;][]{Pick1998}\footnote{The JPL webpage is at \url{https://spec.jpl.nasa.gov}.} are referenced to confirm the transitions and quantum number assignments. In CASSIS, we first divide the molecular species listed in CASSIS into 6 groups with mass numbers in the ranges of $<20$, 20-25, 26-30, 31-35, 36-40 and $>40$. The last group mostly consists of the heaviest and/or most complex species. We further divide the heavy species into 8 sub-groups: those in the vibrational states and those with 1, 2, 3, 4, 5 ,6 and $>6$ atoms. Then, we carefully trawl through the lines of all molecular species in each (sub-)group in the observed frequency ranges to find candidate carriers of the observed line features. Finally, consistency of identified line carriers are examined according to the predicted local thermal equilibrium (LTE) line strengths in CASSIS. In the cases of blended lines, those carriers with apparently much smaller contribution (e.g., minor isotopic or very complex species) than the other are removed. We do not perform further investigations such as excitation analysis to fine tune the line identification, because that will be a subject of quantitative variability study (variation in excitation) for our future papers. The identified molecular lines are listed in Table~\ref{tab:line-list} and labeled in Fig.~\ref{fig: spec_ave}. For clarity, different line features can be labeled in different panels of each sub-plot. However, all blended line carriers of the same line feature are aggregated in the same panel when possible.

The detected line features are attributed to 148 transitions from the following 20 candidate molecules (and/or their isotopic species which are not listed here): C$_2$H, C$_3$H, C$_4$H, c-C$_3$H, c-C$_3$H$_2$, HCN, C$_3$N, HC$_3$N, CH$_2$NH, CH$_3$CN, KCl, KF, KCN, NaCN, NaCl, SiS, SiC, SiCC, SiO, CS and another 81 unidentified line features (229 entries in total in Table~\ref{tab:line-list}). However, the counting of transitions is quite provisional, because more careful analysis and perhaps more multi-transition data are needed to double check whether all the listed molecular transitions are surely detected. Particularly, there is an apparent difficulty in assessing whether those weak narrow line features are independent narrow lines or merely one peak of a weak double-peak feature. 

The strongest lines belong to SiS and HCN in the ground state, reaching 150\,Jy/beam. Both SiS and HCN have relatively strong lines in multiple vibrational states ($v$=0, 1, 2, 3, 4 for SiS and $v_2$=1, 2, 3 for HCN) detected, which are good agents for variability studies. The SiCC and its isotopic lines usually show double peak line profiles that is typical for a spatially resolved envelope, while many hydrocarbon lines such as $^{13}$CCH, C$_3$H, c-C$_3$H, C$_4$H, isotopic C$_4$H lines and C$_4$H lines in vibrational states show negative spectral features around the line center possibly due to the combination of the negative side lobes of the ring like emission features that partially overlap around the phase center, the insufficient image fidelity (sparse sampling of the UV plane), the possible missing of extended smooth emission, and phase noise. Most of the unidentified lines (U-lines) are fainter than 1\,Jy/beam, with only few exceptionally strong ones: U267423, U268663, U268781 and U269427. As will be mentioned in the next subsection, a large number of the U-lines share a common variability feature, which may help constrain their molecular carriers. Many U-lines, as found by \citet[][]{Pate2011} with SMA or \citet[][]{Cern2013} with ALMA, are narrow and thus should be excited in the wind acceleration zone.

Our average spectra can be compared to the observation of \citet{Cern2013}. They observed IRC\,+10216 in similar frequency ranges with a high resolution of $0.6''\times 0.5''$ of the 12\,m-array of ALMA and compared the spectra with their single dish data from IRAM\,30m telescope (beam size of $9.5''$). In the frequency range overlapping with our spectra (268200-268640\,MHz), our spectrum appears more similar to their single dish data because of the similar beam sizes. The two major differences to their work are in line assignments: (1) they assigned our U-line at 268202\,MHz to HCN,$v_2$=6,$v_3$=1,3-2,$\ell$=$2^0$; (2) the three-peak line feature around 268560\,MHz are assigned to a mixture of $^{29}$SiC$_2$,$11_{2,9}$-$10_{2,8}$, KCl,35-34 and $^{13}$C$^{36}$S,6-5 by us, while it was assigned to HCN,$v_2$=4,3-2,$\ell$=$2^e$+$2^f$ by them. However, neither of their high excitation HCN lines is available in the public line list databases. 
%
\begin{figure}[ht!]
\centering
\includegraphics[scale=0.65]{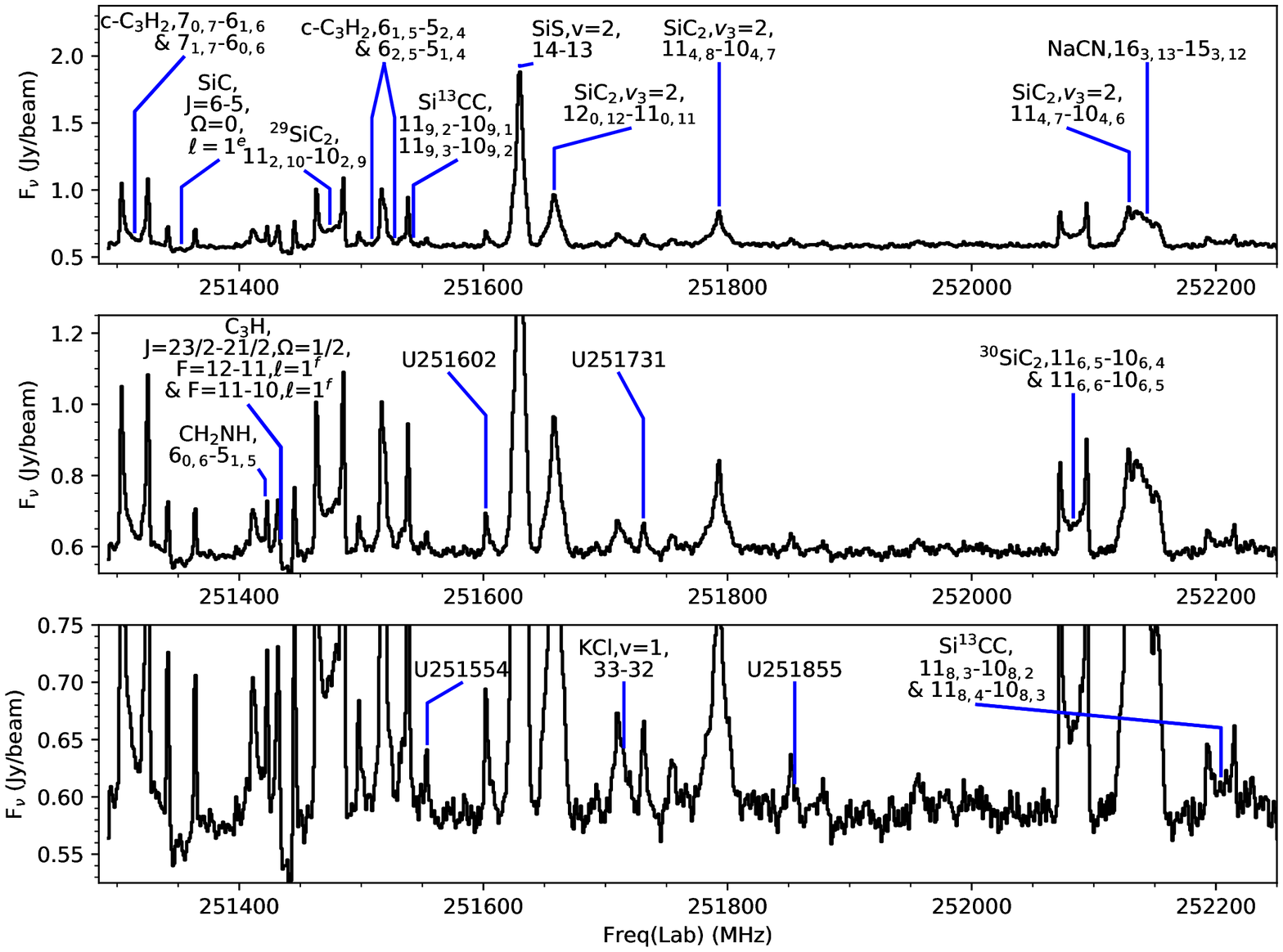}
\includegraphics[scale=0.65]{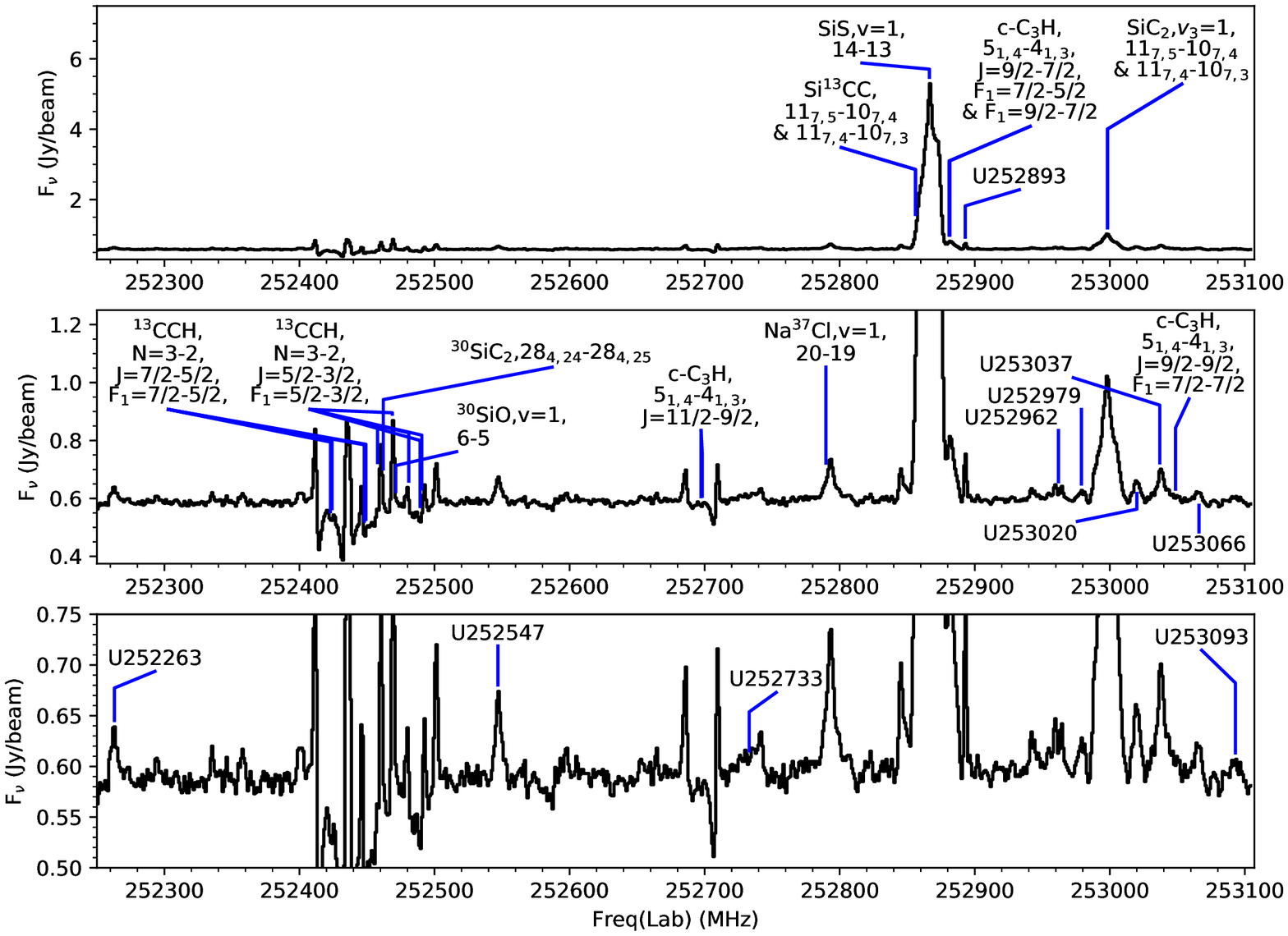}
\caption{Average spectra and spectral line labels. Each sub-figure has three panels that sequentially zoom in to smaller flux ranges from the top panel to bottom panel to show weaker emission features. The lab frequency of each line can be found in Table~\ref{tab:line-list}. \label{fig: spec_ave}}
\end{figure}
\begin{figure}[ht!]
\centering
\addtocounter{figure}{-1}
\includegraphics[scale=0.655]{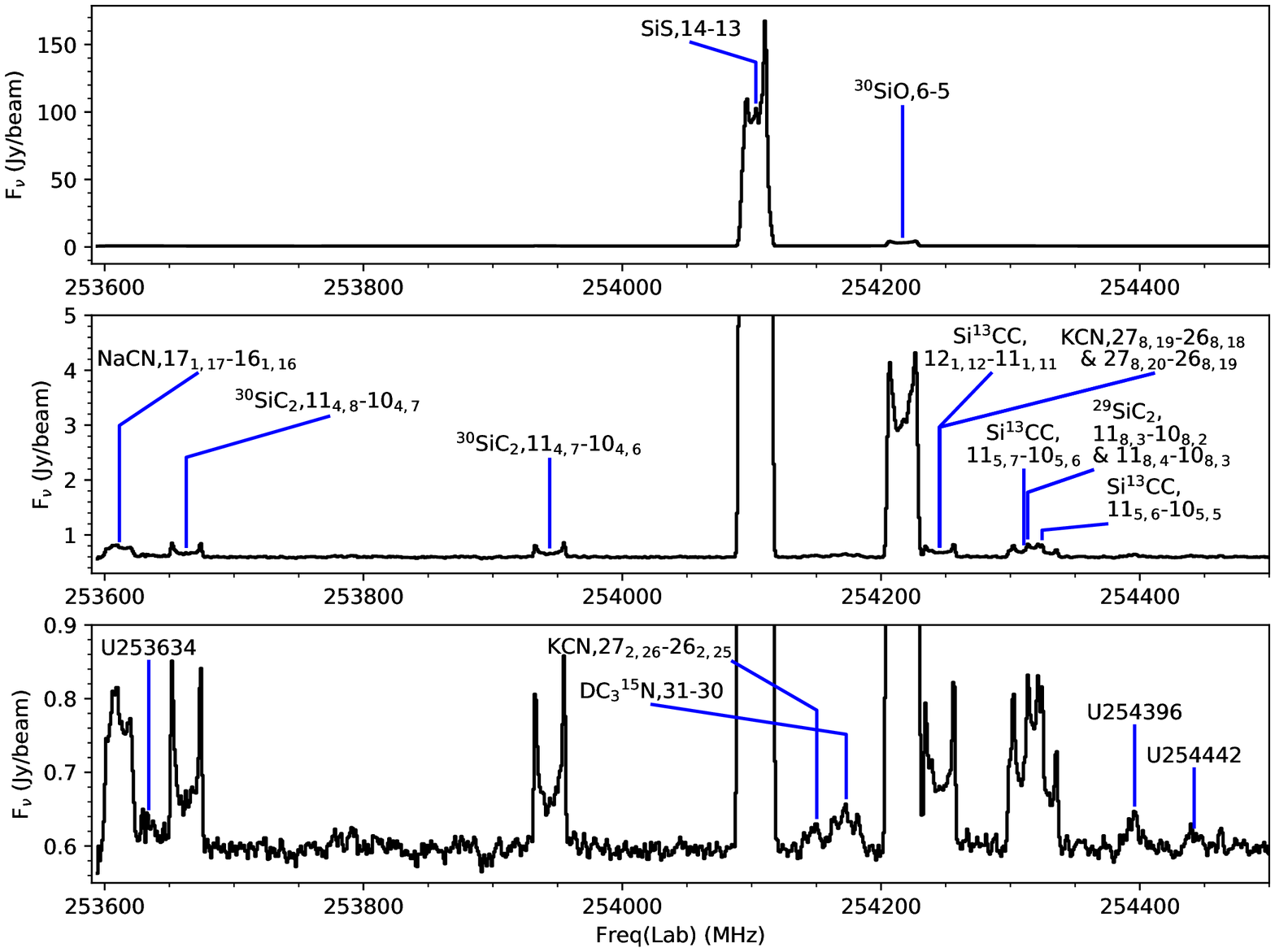}
\includegraphics[scale=0.655]{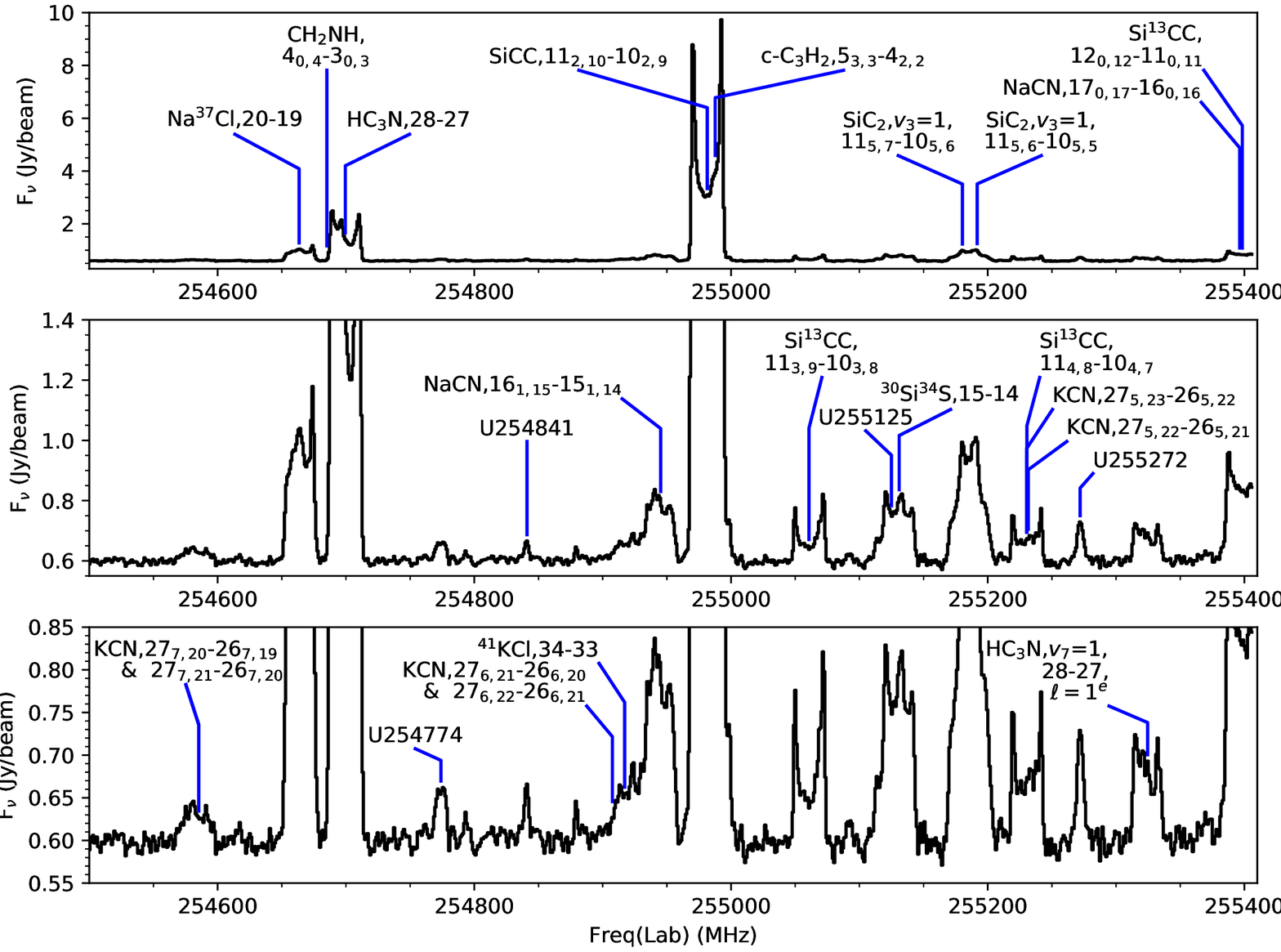}
\caption{(continued)}
\end{figure}
\begin{figure}[ht!]
\centering
\addtocounter{figure}{-1}
\includegraphics[scale=0.655]{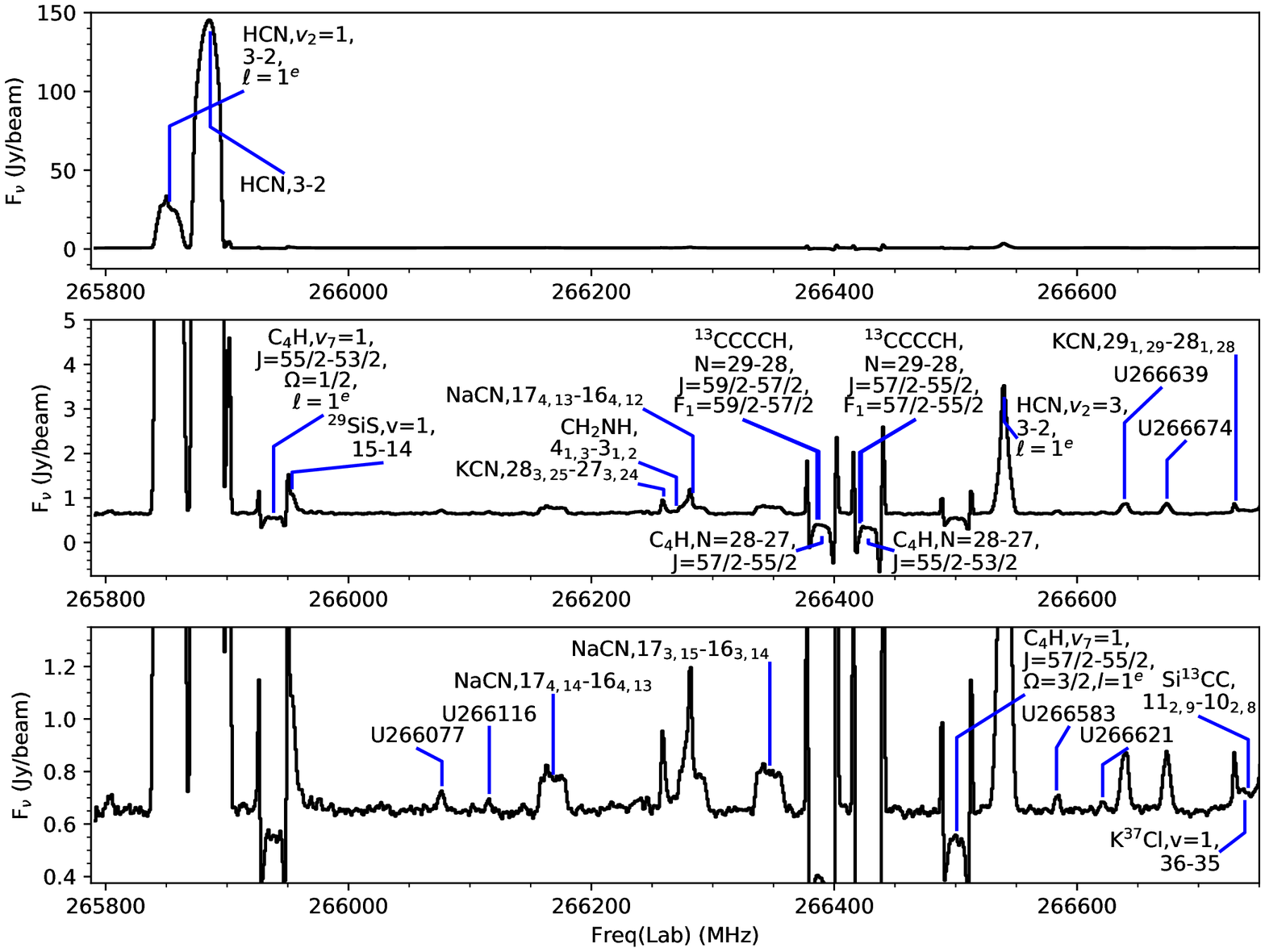}
\includegraphics[scale=0.655]{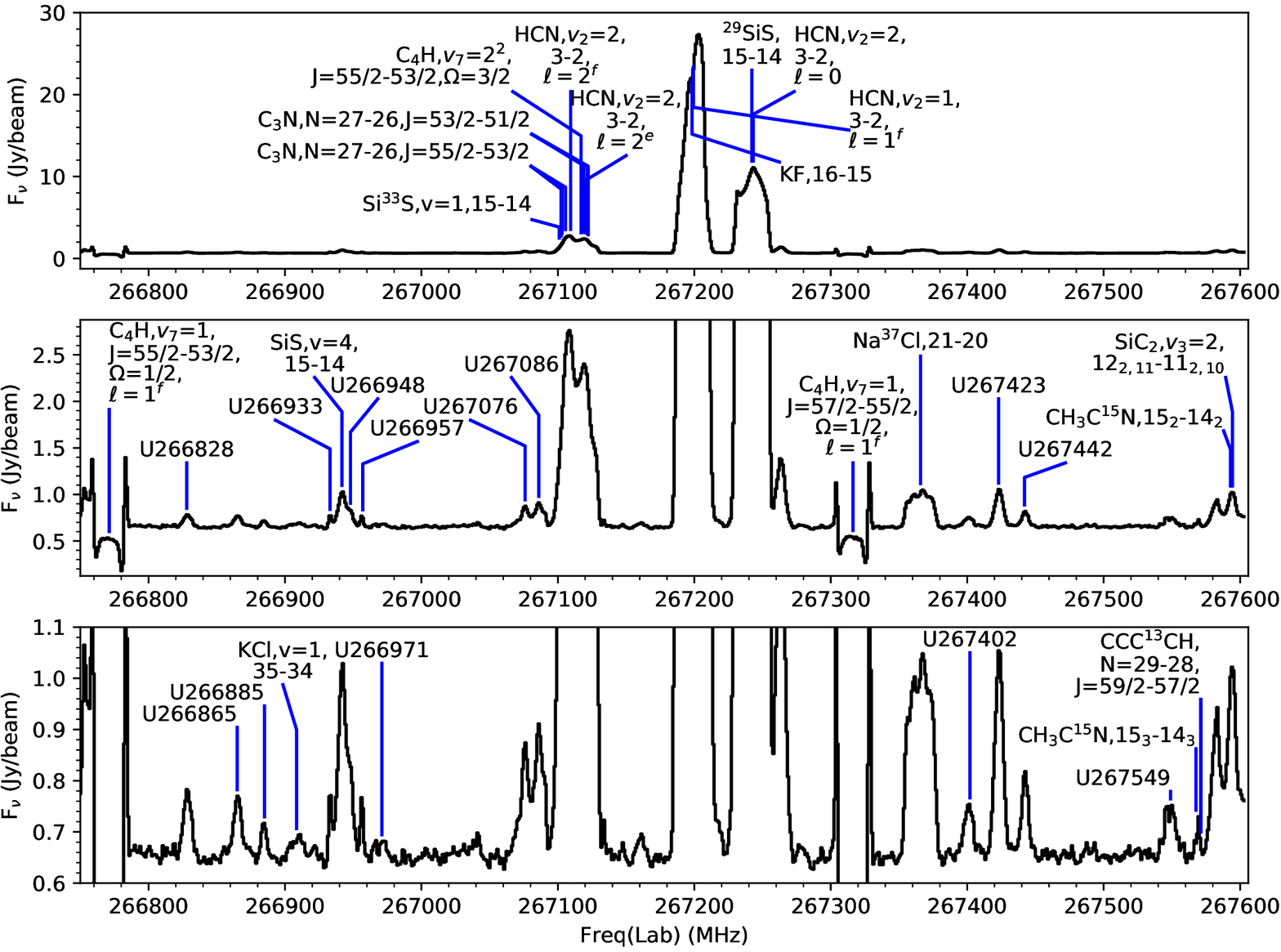}
\caption{(continued)}
\end{figure}
\begin{figure}[ht!]
\centering
\addtocounter{figure}{-1}
\includegraphics[scale=0.655]{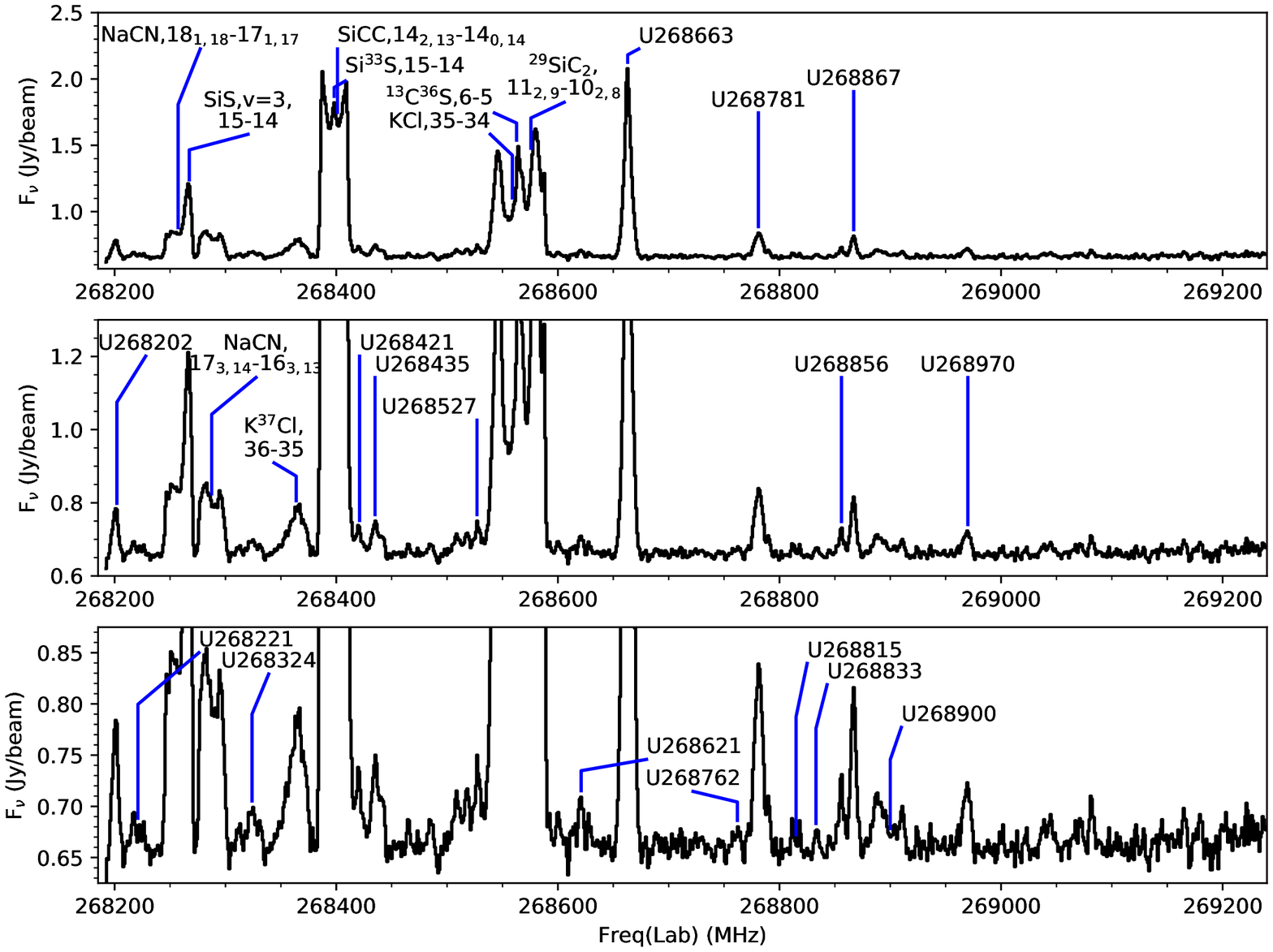}
\includegraphics[scale=0.655]{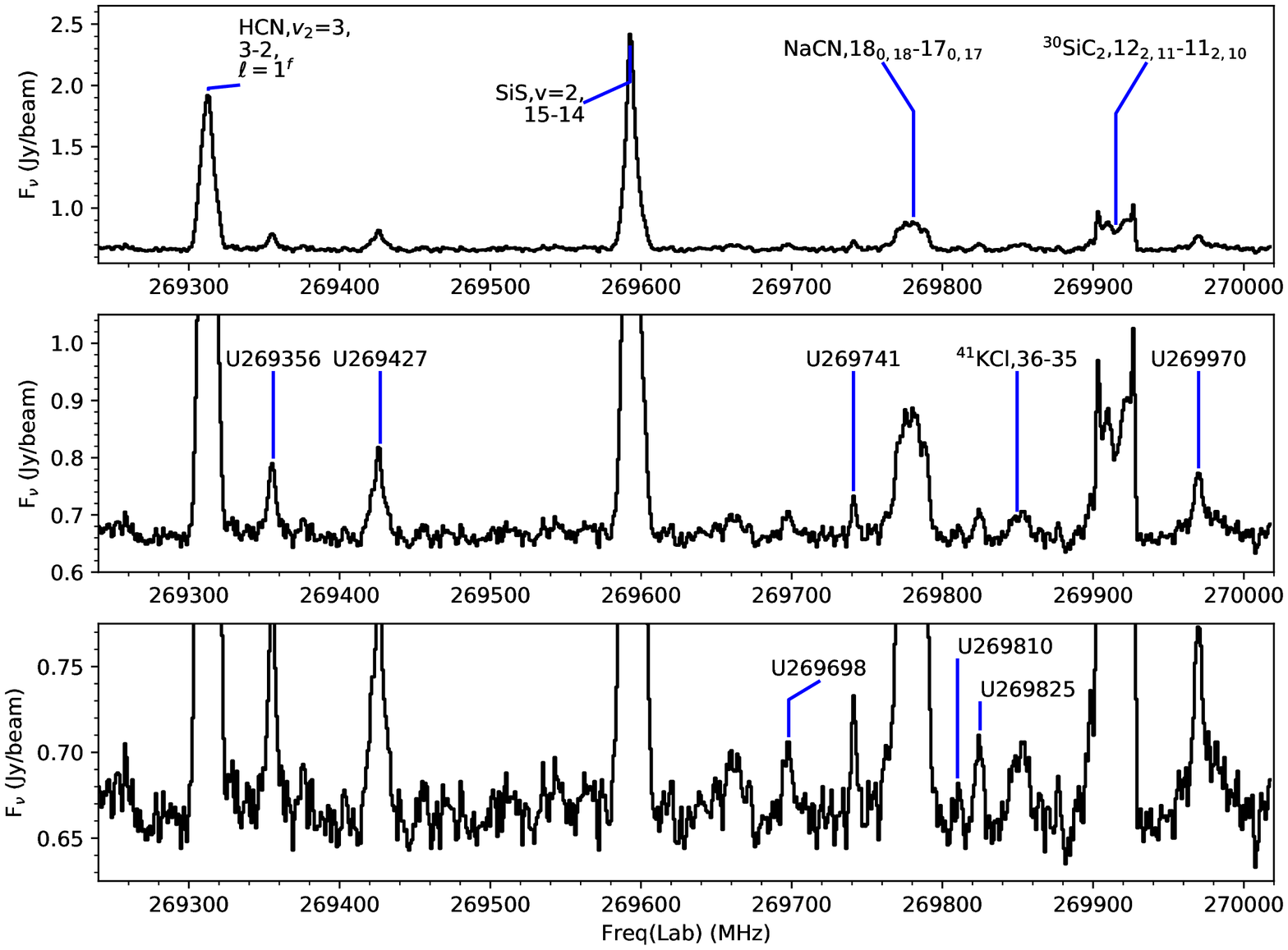}
\caption{(continued)}
\end{figure}
\begin{deluxetable*}{llrllr}[th!]
\tablecaption{List of identified lines from the average spectra. \label{tab:line-list}}
\tablecolumns{6}
\tablenum{2}
\tablewidth{0pt}
\tablehead{
\colhead{Freq(Lab)} &
\colhead{Transition} &
\colhead{$E_{\rm up}$} &
\colhead{Freq(Lab)} &
\colhead{Transition} &
\colhead{$E_{\rm up}$} \\
\colhead{(MHz)} &
\colhead{} &
\colhead{(K)} &
\colhead{(MHz)} &
\colhead{} &
\colhead{(K)} 
}
\startdata
251314.337 & c-C$_3$H$_2$,7$_{0,7}$-6$_{1,6}$ & 50.67 & 252489.308 & $^{13}$CCH,N=3-2,J=5/2-3/2, & 24.28\\ 
251314.343 & c-C$_3$H$_2$,7$_{1,7}$-6$_{0,6}$ & 50.67 &            & \hspace{0.1 in} F$_1$=3/2-1/2,F=3/2-1/2 & \\ 
251352.696 & SiC,J=6-5,$\Omega$=0,$\ell=1^e$ & 158.87 & 252490.623 & $^{13}$CCH,N=3-2,J=5/2-3/2, & 24.28\\ 
251421.265 & CH$_2$NH,6$_{0,6}$-5$_{1,5}$ & 64.07 &            & \hspace{0.1 in} F$_1$=3/2-1/2,F=3/2-3/2 & \\ 
251433.892 & C$_3$H,J=23/2-21/2,$\Omega$=1/2, &  & 252547.0 & U252547 & \\ 
           & \hspace{0.1 in} F=12-11,$\ell$=1$^f$ &  & 252697.373 & c-C$_3$H,5$_{1,4}$-4$_{1,3}$,J=11/2-9/2, & 37.4\\ 
251434.415 & C$_3$H,J=23/2-21/2,$\Omega$=1/2, &  &            & \hspace{0.1 in} F$_1$=11/2-9/2 & \\ 
           & \hspace{0.1 in} F=11-10,$\ell$=1$^f$ &  & 252698.198 & c-C$_3$H,5$_{1,4}$-4$_{1,3}$,J=11/2-9/2, & 37.41\\ 
251474.276 & $^{29}$SiC$_2$,11$_{2,10}$-10$_{2,9}$ & 80.84 &            & \hspace{0.1 in} F$_1$=9/2-7/2 & \\ 
251508.691 & c-C$_3$H$_2$,6$_{1,5}$-5$_{2,4}$ & 47.49 & 252733.0 & U252733 & \\ 
251527.302 & c-C$_3$H$_2$,6$_{2,5}$-5$_{1,4}$ & 47.49 & 252789.82 & Na$^{37}$Cl,v=1,20-19 & 641.54\\ 
251542.682 & Si$^{13}$CC,11$_{9,2}$-10$_{9,1}$ & 224.04 & 252856.152 & Si$^{13}$CC,11$_{7,5}$-10$_{7,4}$ & 164.38\\ 
251542.682 & Si$^{13}$CC,11$_{9,3}$-10$_{9,2}$ & 224.04 & 252856.155 & Si$^{13}$CC,11$_{7,4}$-10$_{7,3}$ & 164.38\\ 
251554.0 & U251554 &  & 252866.468 & SiS,v=1,14-13 & 1162.18\\ 
251602.0 & U251602 &  & 252881.049 & c-C$_3$H,5$_{1,4}$-4$_{1,3}$,J=9/2-7/2, & 37.42\\ 
251629.609 & SiS,v=2,14-13 & 2225.45 &            & \hspace{0.1 in} F$_1$=7/2-5/2 & \\ 
251658.041 & SiC$_2$,$v_3$=2,12$_{0,12}$-11$_{0,11}$ & 588.51 & 252881.59 & c-C$_3$H,5$_{1,4}$-4$_{1,3}$,J=9/2-7/2, & 37.42\\ 
251715.123 & KCl,v=1,33-32 & 604.81 &            & \hspace{0.1 in} F$_1$=9/2-7/2 & \\ 
251731.0 & U251731 &  & 252893.0 & U252893 & \\ 
251754.0 & U251754 &  & 252943.0 & U252943 & \\ 
251793.162 & SiC$_2$,$v_3$=2,11$_{4,8}$-10$_{4,7}$ & 613.75 & 252962.0 & U252962 & \\ 
251855.0 & U251855 &  & 252979.0 & U252979 & \\ 
252083.255 & $^{30}$SiC$_2$,11$_{6,5}$-10$_{6,4}$ & 143.38 & 252998.255 & SiC$_2$,$v_3$=1,11$_{7,5}$-10$_{7,4}$ & 454.64\\ 
252083.255 & $^{30}$SiC$_2$,11$_{6,6}$-10$_{6,5}$ & 143.38 & 252998.257 & SiC$_2$,$v_3$=1,11$_{7,4}$-10$_{7,3}$ & 454.64\\ 
252128.989 & SiC$_2$,$v_3$=2,11$_{4,7}$-10$_{4,6}$ & 613.78 & 253020.0 & U253020 & \\ 
252143.791 & NaCN,16$_{3,13}$-15$_{3,12}$ & 124.05 & 253037.0 & U253037 & \\ 
252204.244 & Si$^{13}$CC,11$_{8,3}$-10$_{8,2}$ & 192.32 & 253048.673 & c-C$_3$H,5$_{1,4}$-4$_{1,3}$,J=9/2-9/2, & 37.42\\ 
252204.244 & Si$^{13}$CC,11$_{8,4}$-10$_{8,3}$ & 192.32 &            & \hspace{0.1 in} F$_1$=7/2-7/2 & \\ 
252263.0 & U252263 &  & 253066.0 & U253066 & \\ 
252347.0 & U252347 &  & 253093.0 & U253093 & \\ 
252422.933 & $^{13}$CCH,N=3-2,J=7/2-5/2, & 24.27 & 253611.375 & NaCN,17$_{1,17}$-16$_{1,16}$ & 112.5\\ 
           & \hspace{0.1 in} F$_1$=7/2-5/2,F=9/2-7/2 &  & 253634.0 & U253634 & \\ 
252424.122 & $^{13}$CCH,N=3-2,J=7/2-5/2, & 24.27 & 253663.045 & $^{30}$SiC$_2$,11$_{4,8}$-10$_{4,7}$ & 104.32\\ 
           & \hspace{0.1 in} F$_1$=7/2-5/2,F=7/2-5/2 &  & 253785.0 & U253785 & \\ 
252447.991 & $^{13}$CCH,N=3-2,J=7/2-5/2, & 24.23 & 253943.918 & $^{30}$SiC$_2$,11$_{4,7}$-10$_{4,6}$ & 104.35\\ 
           & \hspace{0.1 in} F$_1$=5/2-3/2,F=5/2-3/2 &  & 254103.211 & SiS,14-13 & 91.47\\ 
252449.265 & $^{13}$CCH,N=3-2,J=7/2-5/2, & 24.28 & 254150.307 & KCN,27$_{2,26}$-26$_{2,25}$ & 181.56\\ 
           & \hspace{0.1 in} F$_1$=5/2-3/2,F=7/2-5/2 &  & 254173.073 & DC$_3$$^{15}$N,31-30 & 195.19\\ 
252457.865 & $^{13}$CCH,N=3-2,J=5/2-3/2, & 24.28 & 254216.656 & $^{30}$SiO,6-5 & 42.7\\ 
           & \hspace{0.1 in} F$_1$=5/2-3/2,F=7/2-5/2 &  & 254244.965 & Si$^{13}$CC,12$_{1,12}$-11$_{1,11}$ & 81.96\\ 
252462.082 & $^{30}$SiC$_2$,28$_{4,24}$-28$_{4,25}$ & 492.8 & 254245.402 & KCN,27$_{8,19}$-26$_{8,18}$ & 335.16\\ 
252468.774 & $^{13}$CCH,N=3-2,J=5/2-3/2, & 24.28 & 254245.402 & KCN,27$_{8,20}$-26$_{8,19}$ & 335.16\\ 
           & \hspace{0.1 in} F$_1$=5/2-3/2,F=5/2-3/2 &  & 254310.375 & Si$^{13}$CC,11$_{5,7}$-10$_{5,6}$ & 119.79\\ 
252471.372 & $^{30}$SiO,v=1,6-5 & 1790.21 & 254313.374 & $^{29}$SiC$_2$,11$_{8,3}$-10$_{8,2}$ & 198.84\\ 
252480.925 & $^{13}$CCH,N=3-2,J=5/2-3/2, & 24.28 & 254313.374 & $^{29}$SiC$_2$,11$_{8,4}$-10$_{8,3}$ & 198.84\\ 
           & \hspace{0.1 in} F$_1$=3/2-1/2,F=5/2-3/2 &  & 254324.417 & Si$^{13}$CC,11$_{5,6}$-10$_{5,5}$ & 119.79\\ 
\enddata
\end{deluxetable*}

\begin{deluxetable*}{llrllr}[th!]
\tablecaption{(continue.)}
\tablecolumns{6}
\tablenum{2}
\tablewidth{0pt}
\tablehead{
\colhead{Freq(Lab)} &
\colhead{Transition} &
\colhead{$E_{\rm up}$} &
\colhead{Freq(Lab)} &
\colhead{Transition} &
\colhead{$E_{\rm up}$} \\
\colhead{(MHz)} &
\colhead{} &
\colhead{(K)} &
\colhead{(MHz)} &
\colhead{} &
\colhead{(K)} 
}
\startdata
254396.0 & U254396 &  &            & \hspace{0.1 in} F$_1$=59/2-57/2,F=59/2-57/2 & \\ 
254442.0 & U254442 &  & 266385.909 & $^{13}$CCCCH,N=29-28,J=59/2-57/2, & 191.8\\ 
254585.245 & KCN,27$_{7,20}$-26$_{7,19}$ & 296.92 &            & \hspace{0.1 in} F$_1$=59/2-57/2,F=61/2-59/2 & \\ 
254585.245 & KCN,27$_{7,21}$-26$_{7,20}$ & 296.92 & 266387.231 & $^{13}$CCCCH,N=29-28,J=59/2-57/2, & 191.79\\ 
254663.465 & Na$^{37}$Cl,20-19 & 128.4 &            & \hspace{0.1 in} F$_1$=57/2-55/2,F=59/2-57/2 & \\ 
254685.137 & CH$_2$NH,4$_{0,4}$-3$_{0,3}$ & 30.62 & 266387.274 & $^{13}$CCCCH,N=29-28,J=59/2-57/2, & 191.79\\ 
254699.5 & HC$_3$N,28-27 & 177.26 &            & \hspace{0.1 in} F$_1$=57/2-55/2,F=57/2-55/2 & \\ 
254774.0 & U254774 &  & 266389.925 & C$_4$H,N=28-27,J=57/2-55/2, & 185.39\\ 
254793.0 & U254793 &  &            & \hspace{0.1 in} F=28-27 & \\ 
254841.0 & U254841 &  & 266389.926 & C$_4$H,N=28-27,J=57/2-55/2, & 185.39\\ 
254879.0 & U254879 &  &            & \hspace{0.1 in} F=29-28 & \\ 
254907.541 & KCN,27$_{6,22}$-26$_{6,21}$ & 263.77 & 266420.929 & $^{13}$CCCCH,N=29-28,J=57/2-55/2, & 191.85\\ 
254907.558 & KCN,27$_{6,21}$-26$_{6,20}$ & 263.77 &            & \hspace{0.1 in} F$_1$=57/2-55/2,F=57/2-55/2 & \\ 
254917.167 & $^{41}$KCl,34-33 & 214.3 & 266420.973 & $^{13}$CCCCH,N=29-28,J=57/2-55/2, & 191.85\\ 
254945.207 & NaCN,16$_{1,15}$-15$_{1,14}$ & 107.24 &            & \hspace{0.1 in} F$_1$=57/2-55/2,F=59/2-57/2 & \\ 
254981.494 & SiCC,11$_{2,10}$-10$_{2,9}$ & 81.86 & 266422.296 & $^{13}$CCCCH,N=29-28,J=57/2-55/2, & 191.86\\ 
254987.64 & c-C$_3$H$_2$,5$_{3,3}$-4$_{2,2}$ & 41.06 &            & \hspace{0.1 in} F$_1$=55/2-53/2,F=55/2-53/2 & \\ 
255060.768 & Si$^{13}$CC,11$_{3,9}$-10$_{3,8}$ & 90.16 & 266422.296 & $^{13}$CCCCH,N=29-28,J=57/2-55/2, & 191.86\\ 
255092.0 & U255092 &  &            & \hspace{0.1 in} F$_1$=55/2-53/2,F=57/2-55/2 & \\ 
255125.0 & U255125 &  & 266428.184 & C$_4$H,N=28-27,J=55/2-53/2, & 185.45\\ 
255131.019 & $^{30}$Si$^{34}$S,15-14 & 97.97 &            & \hspace{0.1 in} F=27-26 & \\ 
255180.308 & SiC$_2$,$v_3$=1,11$_{5,7}$-10$_{5,6}$ & 406.47 & 266428.184 & C$_4$H,N=28-27,J=55/2-53/2, & 185.45\\ 
255191.739 & SiC$_2$,$v_3$=1,11$_{5,6}$-10$_{5,5}$ & 406.48 &            & \hspace{0.1 in} F=28-27 & \\ 
255230.449 & KCN,27$_{5,23}$-26$_{5,22}$ & 235.72 & 266500.885 & C$_4$H,$v_7$=1,J=57/2-55/2, & 374.86\\ 
255230.503 & Si$^{13}$CC,11$_{4,8}$-10$_{4,7}$ & 103.15 &            & \hspace{0.1 in} $\Omega$=1/2,$\ell=1^e$ & \\ 
255231.586 & KCN,27$_{5,22}$-26$_{5,21}$ & 235.72 & 266540.0 & HCN,$v_2$=3,3-2,$\ell=1^e$ & 3066.4\\ 
255272.0 & U255272 &  & 266583.0 & U266583 & \\ 
255324.556 & HC$_3$N,$v_7$=1,28-27,$\ell=1^e$ & 498.55 & 266621.0 & U266621 & \\ 
255369.0 & U255369 &  & 266639.0 & U266639 & \\ 
255396.367 & NaCN,17$_{0,17}$-16$_{0,16}$ & 112.09 & 266674.0 & U266674 & \\ 
255398.378 & Si$^{13}$CC,12$_{0,12}$-11$_{0,11}$ & 81.85 & 266730.815 & KCN,29$_{1,29}$-28$_{1,28}$ & 195.29\\ 
265803.0 & U265803 &  & 266738.158 & K$^{37}$Cl,v=1,36-35 & 630.66\\ 
265852.709 & HCN,$v_2$=1,3-2,$\ell=1^e$ & 1049.91 & 266740.962 & Si$^{13}$CC,11$_{2,9}$-10$_{2,8}$ & 83.17\\ 
265886.18 & HCN,3-2 & 25.52 & 266771.187 & C$_4$H,$v_7$=1,J=55/2-53/2, & 376.3\\ 
265938.285 & C$_4$H,$v_7$=1,J=55/2-53/2, & 375.73 &            & \hspace{0.1 in} $\Omega$=1/2,$\ell=1^f$ & \\ 
           & \hspace{0.1 in} $\Omega$=1/2,$\ell=1^e$ &  & 266828.0 & U266828 & \\ 
265953.527 & $^{29}$SiS,v=1,15-14 & 1163.44 & 266865.0 & U266865 & \\ 
266077.0 & U266077 &  & 266885.0 & U266885 & \\ 
266116.0 & U266116 &  & 266908.503 & KCl,v=1,35-34 & 630.07\\ 
266168.621 & NaCN,17$_{4,14}$-16$_{4,13}$ & 153.31 & 266933.0 & U266933 & \\ 
266240.0 & U266240 &  & 266941.754 & SiS,v=4,15-14 & 4342.55\\ 
266259.541 & KCN,28$_{3,25}$-27$_{3,24}$ & 207.97 & 266948.0 & U266948 & \\ 
266270.024 & CH$_2$NH,4$_{1,3}$-3$_{1,2}$ & 39.84 & 266957.0 & U266957 & \\ 
266283.634 & NaCN,17$_{4,13}$-16$_{4,12}$ & 153.32 & 266971.0 & U266971 & \\ 
266346.972 & NaCN,17$_{3,15}$-16$_{3,14}$ & 136.61 & 267032.0 & U267032 & \\ 
266385.908 & $^{13}$CCCCH,N=29-28,J=59/2-57/2, & 191.8 & 267076.0 & U267076 & \\ 
\enddata
\end{deluxetable*}

\begin{deluxetable*}{llrllr}[th!]
\tablecaption{(continue.)}
\tablecolumns{6}
\tablenum{2}
\tablewidth{0pt}
\tablehead{
\colhead{Freq(Lab)} &
\colhead{Transition} &
\colhead{$E_{\rm up}$} &
\colhead{Freq(Lab)} &
\colhead{Transition} &
\colhead{$E_{\rm up}$} \\
\colhead{(MHz)} &
\colhead{} &
\colhead{(K)} &
\colhead{(MHz)} &
\colhead{} &
\colhead{(K)} 
}
\startdata
267086.0 & U267086 &  & 267571.105 & CCC$^{13}$CH,N=29-28,J=59/2-57/2, & 192.64\\ 
267101.025 & Si$^{33}$S,v=1,15-14 & 1166.16 &            & \hspace{0.1 in} F$_1$=57/2-55/2,F=59/2-57/2 & \\ 
267102.372 & C$_3$N,N=27-26,J=55/2-53/2, & 179.48 & 267592.943 & CH$_3$C$^{15}$N,15$_{2}$-14$_{2}$ & 131.38\\ 
           & \hspace{0.1 in} F=55/2-55/2 &  & 267594.428 & SiC$_2$,$v_3$=2,12$_{2,11}$-11$_{2,10}$ & 600.48\\ 
267102.87 & C$_3$N,N=27-26,J=55/2-53/2, & 179.48 & 268202.0 & U268202 & \\ 
           & \hspace{0.1 in} F=53/2-51/2 &  & 268221.0 & U268221 & \\ 
267102.87 & C$_3$N,N=27-26,J=55/2-53/2, & 179.48 & 268257.243 & NaCN,18$_{1,18}$-17$_{1,17}$ & 125.37\\ 
           & \hspace{0.1 in} F=55/2-53/2 &  & 268267.357 & SiS,v=3,15-14 & 3294.18\\ 
267102.871 & C$_3$N,N=27-26,J=55/2-53/2, & 179.48 & 268287.569 & NaCN,17$_{3,14}$-16$_{3,13}$ & 136.93\\ 
           & \hspace{0.1 in} F=57/2-55/2 &  & 268324.0 & U268324 & \\ 
267105.562 & C$_3$N,N=27-26,J=55/2-53/2, & 179.48 & 268363.84 & K$^{37}$Cl,36-35 & 238.52\\ 
           & \hspace{0.1 in} F=53/2-53/2 &  & 268398.099 & Si$^{33}$S,15-14 & 103.06\\ 
267109.37 & HCN,$v_2$=2,3-2,$\ell=2^f$ & 2078.12 & 268401.09 & SiCC,14$_{2,13}$-14$_{0,14}$ & 124.94\\ 
267116.983 & C$_4$H,$v_7$=2$^2$,J=55/2-53/2, & 564.66 & 268421.0 & U268421 & \\ 
           & \hspace{0.1 in} $\Omega$=3/2 &  & 268435.0 & U268435 & \\ 
267118.94 & C$_3$N,N=27-26,J=53/2-51/2, & 179.5 & 268464.0 & U268464 & \\ 
           & \hspace{0.1 in} F=53/2-53/2 &  & 268485.0 & U268485 & \\ 
267120.02 & HCN,$v_2$=2,3-2,$\ell=2^e$ & 2078.12 & 268508.0 & U268508 & \\ 
267121.623 & C$_3$N,N=27-26,J=53/2-51/2, & 179.5 & 268518.0 & U268518 & \\ 
           & \hspace{0.1 in} F=51/2-49/2 &  & 268527.0 & U268527 & \\ 
267121.625 & C$_3$N,N=27-26,J=53/2-51/2, & 179.5 & 268558.98 & KCl,35-34 & 232.24\\ 
           & \hspace{0.1 in} F=53/2-51/2 &  & 268562.714 & $^{13}$C$^{36}$S,6-5 & 45.11\\ 
267121.625 & C$_3$N,N=27-26,J=53/2-51/2, & 179.5 & 268575.684 & $^{29}$SiC$_2$,11$_{2,9}$-10$_{2,8}$ & 84.04\\ 
           & \hspace{0.1 in} F=55/2-53/2 &  & 268621.0 & U268621 & \\ 
267122.209 & C$_3$N,N=27-26,J=53/2-51/2, & 179.5 & 268663.0 & U268663 & \\ 
           & \hspace{0.1 in} F=51/2-51/2 &  & 268762.0 & U268762 & \\ 
267147.0 & U267147 &  & 268781.0 & U268781 & \\ 
267161.0 & U267161 &  & 268815.0 & U268815 & \\ 
267198.324 & KF,16-15 & 109.05 & 268833.0 & U268833 & \\ 
267199.283 & HCN,$v_2$=1,3-2,$\ell=1^f$ & 1050.04 & 268856.0 & U268856 & \\ 
267242.218 & $^{29}$SiS,15-14 & 102.62 & 268867.0 & U268867 & \\ 
267243.15 & HCN,$v_2$=2,3-2,$\ell=0$ & 2056.38 & 268900.0 & U268900 & \\ 
267316.334 & C$_4$H,$v_7$=1,J=57/2-55/2, & 375.4 & 268970.0 & U268970 & \\ 
           & \hspace{0.1 in} $\Omega$=1/2,$\ell=1^f$ &  & 269312.89 & HCN,$v_2$=3,3-2,$\ell=1^f$ & 3066.67\\ 
267365.833 & Na$^{37}$Cl,21-20 & 141.23 & 269356.0 & U269356 & \\ 
267402.0 & U267402 &  & 269427.0 & U269427 & \\ 
267423.0 & U267423 &  & 269592.744 & SiS,v=2,15-14 & 2238.39\\ 
267442.0 & U267442 &  & 269662.0 & U269662 & \\ 
267549.0 & U267549 &  & 269698.0 & U269698 & \\ 
267567.678 & CH$_3$C$^{15}$N,15$_{3}$-14$_{3}$ & 167.16 & 269741.0 & U269741 & \\ 
267571.101 & CCC$^{13}$CH,N=29-28,J=59/2-57/2, & 192.64 & 269780.824 & NaCN,18$_{0,18}$-17$_{0,17}$ & 125.04\\ 
           & \hspace{0.1 in} F$_1$=59/2-57/2,F=59/2-57/2 &  & 269810.0 & U269810 & \\ 
267571.101 & CCC$^{13}$CH,N=29-28,J=59/2-57/2, & 192.64 & 269825.0 & U269825 & \\ 
           & \hspace{0.1 in} F$_1$=59/2-57/2,F=61/2-59/2 &  & 269849.558 & $^{41}$KCl,36-35 & 239.84\\ 
267571.104 & CCC$^{13}$CH,N=29-28,J=59/2-57/2, & 192.64 & 269915.085 & $^{30}$SiC$_2$,12$_{2,11}$-11$_{2,10}$ & 92.85\\ 
           & \hspace{0.1 in} F$_1$=57/2-55/2,F=57/2-55/2 &  & 269970.0 & U269970 & \\ 
\enddata
\end{deluxetable*}

\subsection{Sample light curves} \label{subsec:example-lightcurves}

\begin{figure}[ht!]
\centering
\includegraphics[scale=0.6]{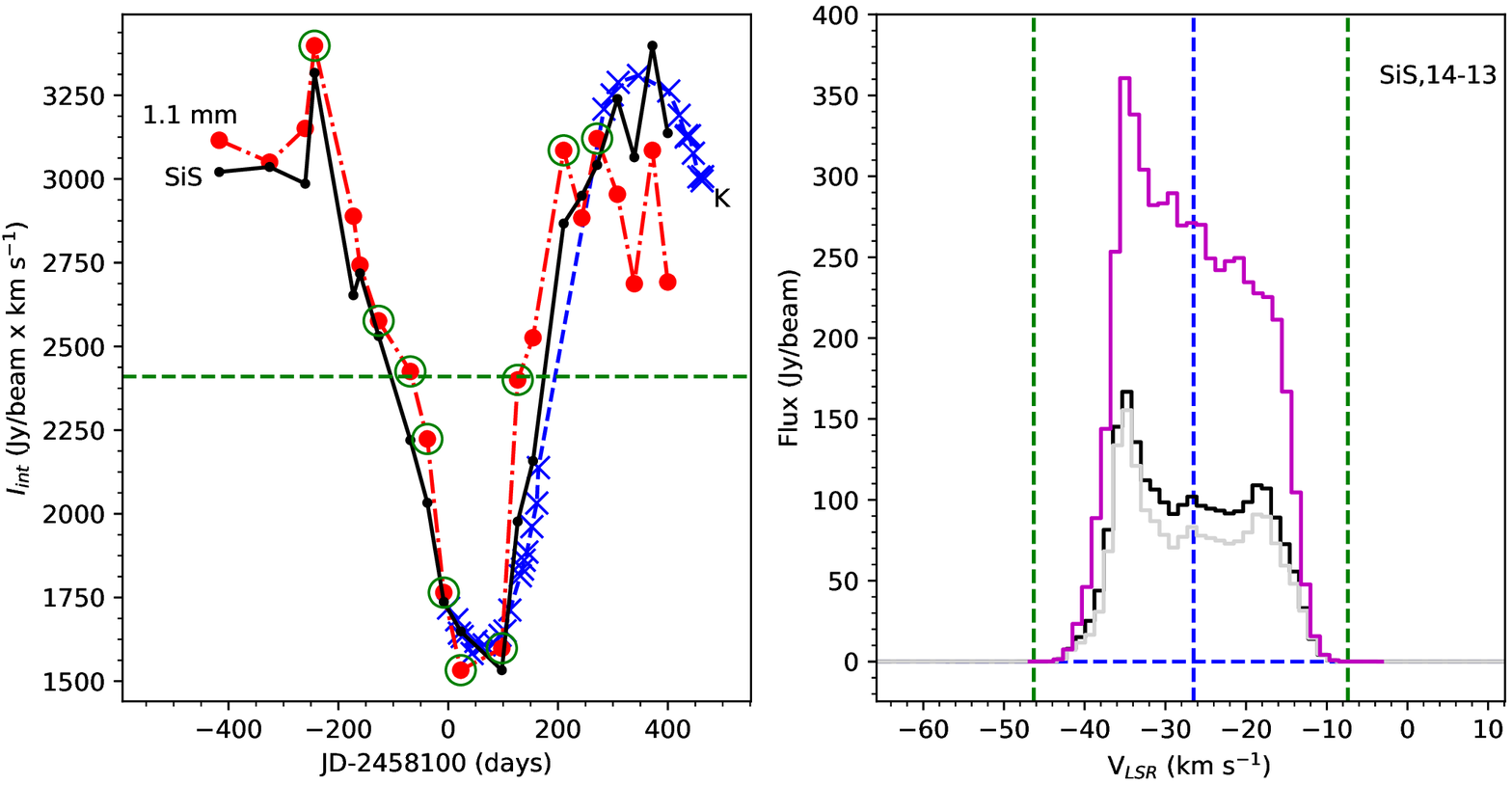}
\caption{({\em Left}) Sample light curves of 1.1\,mm continuum flux (red dots and dash dotted line), near infrared {\em K}-band strength (in negative magnitude; blue crosses and dashed line), and integrated line strength of SiS,14-13 (black dots and solid line). The former two have been normalized to have the same extrema as the SiS line light curve to falicitate comparison. 
Error bars (not shown) are discussed in Sect.~\ref{subsec:example-lightcurves}. The green horizontal dashed line divides all the ALMA monitoring epochs into two groups from which the averaged maximum- and minimum-phase spectra are computed for Fig.~\ref{fig: compare-max-min1}, while  the large green circles indicate the selected ALMA monitoring epochs that are used to compute the average line profiles in that figure.
({\em Right}) The average line profiles of the SiS,14-13 line obtained with the synthesized beam of $4''\times 8''$ (black) and a $29''$ circular Gaussian beam (light gray) convolved to the ALMA map and our single dish monitoring data \citep[magenta, with a $29''$ beam;][]{He2017}. The baseline and systemic velocity (-26.5\,km\,s$^{-1}$) are marked with blue dashed lines. The velocity range to compute the integrated flux of the light curve in the left panel is indicated with the green vertical lines.} \label{fig: SiS lightcurve}
\end{figure}

We first show sample light curves of the 1.1\,mm continuum, the integrated flux of the SiS,14-13 line and the {\em K}-band NIR flux (in negative magnitude) in the left panel of Fig.~\ref{fig: SiS lightcurve}. Both the continuum and {\em K}-band light curves have been normalized to have the same maximum and minimum values as the SiS,14-13 line light curve to facilitate comparison of their morphology. The 1.1\,mm continuum fluxes are obtained from the ALMA data cubes in the following way. After the {\em cleaning} of all frequency channels and the extraction of the spectra from one synthesized beam, the line free frequency channels in the extracted spectra of all four spectral windows are collected and fit by a single power law function $F_\nu=F_{\nu_0}(\nu/\nu_0)^\beta$ with $\nu_0$=260655\,MHz for each epoch; shown in the figure is just the fiducial flux $F_{\nu_0}$. The {\em K}-band light curve is constructed with negative magnitudes which are more appropriate than fluxes to express the very large dynamical range of variation. The error bars of the NIR photometry are very small ($<1\%$). The uncertainty of the ALMA data (continuum and lines) is dominated by flux calibration uncertainty which we take as $8\%$ here. However, as we described in Sect.~\ref{sec:data reduction}, the uncertainty of the relative variation between the 1.1\,mm continuum and the SiS,14-13 light curves should be immune of the flux calibration error, but be dominated by other uncertainties (e.g., pixel flux fluctuations and baseline fitting errors) that have not been well constrained. Therefore, we do not display the error bars in this figure, but discuss the uncertainties in future work. 

It is clear from the left panel of Fig.~\ref{fig: SiS lightcurve} that all three observed light curves roughly agree to each other, particularly with well-matching times of minima. However, clear discrepancies can be  seen around the second maximum phase where the 1.1\,mm flux seems to rise up faster than the SiS-line and {\em K}-band light curves and stop the rising earlier as well. Then, the 1.1\,mm flux starts to decline with significant oscillations. The discrepancies between the 1.1\,mm continuum and the SiS line amount up to about $12\%$ around the NIR maximum time, while their relative uncertainties should be much smaller than the nominal flux calibration uncertainties of $8\%$, thus this discrepancy is significant and must be due to physical processes in the target. Our data also cover part of the previous maximum time and a quirky feature, a sharp peak at the 4th epoch, also appears there. Unfortunately, because the monitoring cadence was lower by a factor of two and the execution was not very homogeneous, the oscillation behavior can not be confirmed there. The nice agreement between the simultaneously monitored NIR and SiS-line light curves in Fig.~\ref{fig: SiS lightcurve} hints that the phase shift of 0.2 as found in the single dish data (beam size of $25''$) by \citet[][]{Fonf2018,Pard2018} is not seen in our ACA data (with a beam of $4''\times 8''$). This is confirmed by all other strong lines involved in this work (light curves not shown). The reason for this difference is unclear, but could be related to both the smaller beam and missing flux (discussed in the next paragraph) of our ACA data. 


The right panel of Fig.~\ref{fig: SiS lightcurve} shows three versions of the average line profile of SiS,14-13: the one from our single dish work \citep[magenta curve, with a nominal efficiency of 46.6\,Jy/K for a 10\,m telescope;][]{He2017}, the one extracted from the ACA line maps reduced in the way described in Sect.~\ref{sec:data reduction} (black curve), and the one extracted similarly but from the ACA maps smoothed (using {\em imsmooth}) by a large round Gaussian beam of $29''$ that is comparable to the single dish beam (light gray curve). The baseline of the ACA spectra is removed by the procedure described in next section. The SiS-line light curves in the left panel of the figure is just integrated (on the black curve) between the two velocities marked by the green vertical dashed lines that engulf all channels with significant emission. The average ACA line profile obtained with the larger beam (light gray curve) is counter-intuitively slightly weaker than that obtained with the original smaller synthesized beam (black curve). This is, on one hand, because a strong extended smooth emission component is missing in the ACA data due to the lack of short interferometric baselines, which can be easily inferred from the big differences between the ACA and single dish spectra in the figure, and on the other hand, also because the larger beam of $29''$ includes a significant contribution from the negative bowls that are located at radii of about $9''$ from the phase center. The SiS line profiles from the ACA data show double peak shapes, with the two peaks at velocity shifts of around $\pm8.5$\,km\,s$^{-1}$ with respect to the stellar velocity of -26.5\,km\,s$^{-1}$. The position of the blue-shifted peak is identical to that of the average line profile (the magenta curve) and the blue-shifted maximum-variation channel in our single dish work \citep[see in Fig.~3 and Sect.~3.3.1 of][]{He2017}, while the red-shifted peak is closer to the systemic velocity than the red-shifted maximum-variation channel in the single dish work (+10.3\,km\,s$^{-1}$). Therefore, the two peak positions are more symmetric with respect to the systemic velocity in the ALMA data. 
A more quantitative comparison with the single dish line-profile variations will be performed in future.

\subsection{Millimeter line variations} \label{subsec:line-var}

This subsection gives a qualitative overview of the millimeter line variations uncovered by our ALMA monitoring. Due to the reasons stated in Sect.~\ref{sec:data reduction}, quantitative analysis will be left for the new interferometric imaging approach in future work. 

As proposed in our single dish work \citep{He2017}, it is possible to apply the so-called `in-band calibration' approach to at least partly remove the dominant flux calibration uncertainties. As we will see below, we indeed have such stable lines for this purpose. However, because the light curves of these tentative stable lines still need to be double checked through quantitative analysis, we also postpone this operation to future work when the new imaging approach becomes available. Thus, only spectral data without the in-band calibration are used in this work.

In order to trace flux variation in spectral lines, we first remove the underlying spectral baseline (the continuum). Although it is possible to fit a baseline to a whole spectral window or even all four spectral windows, we mainly perform baseline fitting to line free channels in the vicinity of each line to enable the comparison of the pros and cons of the different baseline-fitting approaches in future quantitative analysis. As can be seen in Fig.~\ref{fig: spec_ave}, some spectral line features appear side by side or even blended with each other. They can be treated as a single group of line features (hereafter line groups) that can share the same spectral baseline. In total, 125 such line groups are defined in our data. We fit and remove the continuum in the image domain for each line group individually. For each line group, the minimum frequency width that encloses all the emission channels of the member lines is defined as its frequency width. The frequency channels between neighboring line groups probe continuum and weak undetected lines (taken as part of continuum in this work). In order to reliably define the spectral baseline, it has been guaranteed that the total width of continuum channels on both side of each line group must be no less than its own frequency width. In few exceptional cases when a line group appears at the edge of a spectral window so that the edge side may not have enough continuum channels, the continuum channels at the other side is increased to guarantee a total length of continuum channels being no less than twice of the frequency width of the line group. Finally, a linear baseline is fit to the continuum channels of each line group and subtracted at each epoch. 

In order to qualitatively demonstrate the line profile variations, we divide the 21 observed epochs into two groups: those in the maximum and those in the minimum halves of the 1.1\,mm continuum light curve. The dividing line between the two groups, as shown by the green horizontal dashed line in the left panel of Fig.~\ref{fig: SiS lightcurve}, is empirically set at $2410$\,Jy/beam\,km\,s$^{-1}$ (the unit is for the integrated line flux of SiS because the continuum light curve has been normalized to it in the figure). The 16 epochs above this line is averaged together to make the maximum-phase line profiles, while the 5 epochs below the line make the minimum-phase line profiles. Finally, in order to avoid being biased by too many epochs in the maximum-phase, only 5 epochs above the horizontal line are selected to average with the 5 minimum-phase epochs to make the average line profiles (the 10 selected epochs are marked out by large green circles in the left panel of Fig.~\ref{fig: SiS lightcurve}). We stress that the 10 selected epochs are roughly symmetrically distributed only on the 1.1\,mm continuum light curve, so that the average spectrum of a line feature will appear at the middle of its maximum- and minimum-phase spectra only when the line feature varies very closely in phase or in anti-phase with the continuum. Therefore, a significant bias of the average spectrum toward any one of the other two spectra is an indication of a different variation mode than the correlation or anti-correlation with the continuum.

The baseline-free maximum-, minimum-phase and average line profiles of all the 125 line groups are shown with red, blue and black colors respectively in Fig.~\ref{fig: compare-max-min1}. A virtue of this presentation is that the unknown flux uncertainties have been suppressed to some degree by averaging. The relative strengths of the three line profiles are different in different line groups. We try below to classify them by eyes from the plots. Because we have plotted all line profiles in a uniform manner, a discernible profile difference by eyes roughly corresponds to $>$5-10\% of the line-peak strengths. For very weak lines, such a level of variation may be overwhelmed by baseline noise (the channel-to-channel fluctuations). Although we have adopted a nominal flux calibration uncertainty of $\sim$8\% for a single observation, the averaging of multiple epochs, 16, 5 and 10 epochs for the maximum-, minimum-phase and average line profiles, can reduce it to about 2\%, 3.6\%, 2.5\%, respectively, which makes the differentiation of the aforementioned $>$5\% flux variation possible. Furthermore, a profile variation is identified only when it occurs in at least about $1/3$ channels of a line profile. The main findings are as follows.
\begin{itemize}
    \item Stable lines
    
    The stable lines are very important, because they can serve as calibrators to partly remove the flux calibration error through the so called {\em in-band calibration} method advocated in our single dish paper \citep{He2017}. It is found that several species that are expected to have extended spatial distribution do not show significant variation between maximum- and minimum-phase spectra in Fig.~\ref{fig: compare-max-min1}, for example, c-C$_3$H$_2$ (251314, 251518\,MHz), CH$_2$NH (251421, 254685, 266270\,MHz), SiCC (254981, 268401\,MHz), $^{29}$SiCC (251474, 254313\,MHz), $^{30}$SiCC (252183, 252462, 253663, 253944, 269915\,MHz), Si$^{13}$CC (251543, 254245, 254324, 255061, 255230, 266741\,MHz), NaCN (252144, 253611, 254945, 266169, 266284, 266347, 268257, 269781\,MHz), KCN (254908, 266730\,MHz), HC$_3$N (254700\,MHz), Na$^{37}$Cl (254663\,MHz), $^{30}$Si$^{34}$S (255131\,MHz) and K$^{37}$Cl (266738\,MHz). Only few U-lines show little variation in the figure; they are U251554, U251602, U254793, U255125 and U265803. The figure panels of the lines can be found by their frequencies. Unfortunately, the lines of C$_4$H in both ground and vibrational states, which were among the major calibrator lines in our single dish work, all show the negative artifacts in our data, as mentioned in Sect.~\ref{subsec:ave_spec_line_identification}. The upper level energies $E_{\rm up}$ of these stable lines range from about 30\,K (for the c-C$_3$H$_2$ and CH$_2$NH lines) to nearly 500\,K (for some SiCC lines). Usually the lines from heavy species such as SiCC, NaCN, KCN and their isotopolgues have higher $E_{\rm up}$. 
    
    \item Lines varying in phase with NIR light
    
    Most of the other strongest lines vary in phase with the NIR light, i.e., the maximum-phase spectrum (the line profile in  red color) is above the minimum-phase one (blue color). This includes all the SiS,14-13 lines in the vibrational states of $v$=0, 1, 2, 3, 4 (254103, 252866, 251630, 268267, 266942\,MHz), all the HCN,3-2 lines in the $v_2$=0, 1, 2, 3 states (265886, 265853, 267199, 267243, 267109, 267120, 266540, 269313\,MHz), and some lines of Si$^{33}$S (268398\,MHz), $^{30}$SiO (254217\,MHz), Na$^{37}$Cl (267366\,MHz), KCl (268559\,MHz), K$^{37}$Cl (268364\,MHz), and some relatively strong U-lines such as U266077, U266639, U266674, U266828, U266865, U267423, U268202, U268663, U268867, U269741, U269825 and U269970. The upper level energies of the ground-state varying transitions can be as low as $E_{\rm up}$=26\,K (HCN\,3-2) or 43\,K ($^{30}$SiO), while the other identified lines all belong to relatively heavier species and have upper level energies around 100-200\,K. Thus, there is no distinction in $E_{\rm up}$ values between the varying and stable lines. This is also true for the anti-correlation lines in the next paragraph. Therefore, collision dominated excitation of the pure rotational levels is not supported by these results to be responsible for the observed line variations. There are evidences (to be detailed in future papers) that the lines in vibrational states vary  in a similar manner as their corresponding ground-state lines, which strongly hints on the importance of IR excitation in the variation driver of at least some lines. 
    
    \item Lines varying in anti-correlation with NIR light
    
    An interesting finding of this work is that several lines in the vibrational states, e.g. the lines of SiCC,$v_3$=1, 2 (251128, 251658, 251793, 252998, 255191, 267594\,GHz), Na$^{37}$Cl,$v$=1 (252790\,MHz) and KCl,$v$=1 (251715\,MHz) are stronger in the minimum NIR phase than in the maximum phase, while the detected ground-state lines of the same molecules belong to the stable group (see above). Equally interesting is that many relatively strong U-lines also show this behavior, e.g., U251731, U251754, U252263, U252547, U253020, U253037, U254774, U254841, U255272, U254879, U266583, U266885, U267402, U267442, U268464, U268485, U268781, U268833, U268856, U268970, U269356 and U269427. A salient feature of these lines is that they all have a single peak line shape. We speculate that this variability behavior could be due to energy level competition in the IR excitation, similar to the case of HNC lines modeled by \citet{Cern2014}. 
    
    \item Symmetry with respect to the average spectrum
    
    Most of the lines that show significant differences between the maximum- and minimum-phase spectra also have the average spectrum roughly of intermediate strength between the two. Because the epochs selected to construct the average spectra have been balanced between the two variation phases on the 1.1\,mm-continuum light curve, this symmetry indicates that the variation of most of the lines is either correlated or anti-correlated with the continuum variation. However, there are still several exceptional cases in which the average spectrum is closer to the minimum-phase spectrum (never the opposite, however), which signifies that these lines are neither correlated nor anti-correlated with the continuum. For example, this is the case for the following lines that vary in phase with the NIR light: SiS,$v$=4,15-14 (266941.754\,MHz), U266639, U266674, U267423, and U268020; another group of examples that vary in anti-correlation with NIR light are SiCC,$v_3$=2,$12_{0,12}$-$11_{0,11}$ (251658.041\,MHz), Na$^{37}$Cl,$v$=1,20-19 (252789.82\,MHz), U253020, U267402, U268464, U268485 and U268833. 

    \item Comparison to Single dish results
    
    Eight light curves were studied in detail in our single dish work \citep{He2017}, but only six of them can be compared to the ACA results, because the other two lines (of C$_4$H) suffer from the interferometric baseline problems in the ACA data, showing negative artifacts, as mentioned earlier. Two of the six line features (SiS\,14-13 at 254103\,MHz and the blended lines of C$_3$N+HCN+C$_4$H at 267120\,MHz) follow the NIR light curve well in both the single dish and ACA data. However, for the rest four line features that were anti-correlated with the NIR light in the single dish results, none of them shows the anti-correlation in the ACA data: two of them (HCN,$v_2$=1,3-2,$\ell$=$1^f$ at 267199\,MHz and the blended lines of $^{29}$SiS+HCN,$v_2$=$2^0$ at 267243\,MHz) show strong variations in phase with the NIR light and continuum; $^{30}$SiO,6-5 at 254216\,MHz varies similarly but with a smaller amplitude; the blended line feature of Na$^{37}$Cl+CH$_2$NH+HC$_3$N at 254700\,MHz shows modest variation. 
    
    Why are the anti-correlation cases in the single dish data missing in the ACA results, while the correlated cases remain unchanged? The ALMA monitoring started about 7 yrs ($\sim 4$ stellar pulsation cycles) later than our single dish work and thus secular changes of stellar pulsations could have occurred. However, the different phase behaviors of the different lines cannot be explained solely by this; some unknown mechanisms operating in the circumstellar envelope (CSE) are necessary to bring differences among them. The mechanisms are possibly related to the missing flux in the ACA data, as discussed earlier, and the different telescope beam sizes involved in the two works: $29''$ for the single dish vs $4''\times 8''$ (or $6''$) for ACA. However, radial phase delay due to radiation transportation is unlikely, because the difference in beam radii only corresponds to $\sim1552$\,AU at a source distance of 135\,pc, which can produce a phase lag of only $\sim9$ days with the speed of light. This is insufficient to explain the observed change from anti-correlations to correlations with the NIR or mm continuum light. The only known candidate mechanism is that proposed by \citet{Cern2014} from their simulations in which the evacuation of molecular populations in low $J$ energy levels by IR excitation is different for different molecular lines at different radii of the CSE. We will analyze the time series of ALMA ACA maps and do simulations to further explore this idea in coming papers. 

    \item Remarks on special species
    
    Na$^{37}$Cl is a special case. Its J=20-19 (254663.465\,MHz) and 21-20 (267365.833\,MHz) transitions have similarly strong broad line profiles, with peak fluxes of about 1.5\,Jy/beam. However, while the former shows little difference between its maximum- and minimum-phase line profiles in Fig.~\ref{fig: compare-max-min1}, the latter has its line profile definitely varying in phase with the NIR light. Another line of it in the vibrational state, Na$^{37}$Cl,$v$=1,J=20-19, is tentatively identified around 252789.820\,MHz, but shows a frequency offset of about -3\,MHz w.r.t. the observed narrow emission peak. This tentative vibrational line varies in anti-correlation with the NIR light. 
    
    K$^{37}$Cl is detected in its J=36-35 transition in both the ground (268363.84\,MHz) and the $v$=1 vibrational (266738.158\,MHz) states. While the ground state line shows slightly stronger line in the NIR maximum phase, the vibrational line does not show significant variation between the NIR minimum- and maximum-phase. This is not expected if IR excitation would be the driver of the variation in the ground state.
    
    The SiC,J=6-5,$\Omega$=0,$\ell$=1$^e$ line is only tentatively assigned for the double peak line profile around 251352.696\,MHz. Its double peaks are varying in-phase with the NIR light whilst it seems to possess a faint central peak that is varying in an opposite sense.
    

\end{itemize}



\section{Summary} \label{sec: summary}

We report the first results of an on-going monitoring of 1.1\,mm continuum and line variations of IRC\,+10216 using ALMA ACA 7-m array which has covered about 817 days. A NIR photometric monitoring was also initiated. 

Qualitative presentations of all line features are presented and very rich lines with diverse line profile shapes and variability behaviors have been uncovered. Although a handful of lines appear to be stable during our monitoring and some lines vary in anti-correlation to the NIR light, most strong lines vary roughly in-phase with the NIR light and mm continuum. The variation behaviors of lines in vibrational ground state are not correlated with the upper level energies, indicating that collisional excitation is not responsible for the observed variations. The narrow line widths of many varying lines indicate their origin to be in the hot dense inner CSE, making them potentially good probes to the wind launching dynamics and energetics.

Although the variations of some molecular lines are found to be consistent between the ALMA and existing single dish monitoring of the same star, tricky differences have also been uncovered. The anti-correlations between the light curves of several line features with NIR light in single dish data are not seen in the ALMA data, nor is the phase shift of 0.2 as claimed in a literature. The differences must be related to the smaller beam size and missing flux of the ALMA data. But, a mechanism that operates in the CSE, such as the differential evacuation of the rotational levels by IR excitation, is still necessary for a viable interpretation.

Very rich information in the detailed light curves of the continuum, line intensity and line shapes are yet to be explored in our future work when our novel interferometric imaging approach would become available for a more statistical treatment of the flux uncertainties. 

\acknowledgments

We thank the anonymous referee for the very useful discussions. J.H.He thanks the National Natural Science Foundation of China under Grant Nos. 11873086 and U1631237 and the support by Yunnan Province of China (No.2017HC018). R.E.M. gratefully acknowledges support by VRID-Enlace 218.016.004-1.0, FONDECYT 1190621, and the Chilean Centro de Excelencia en Astrof{\'{i}}sica y Tecnolog{\'{i}}as Afines (CATA) BASAL grant AFB-170002. M.Sch and R.Sz acknowledge support from the grant 2016/21/B/ST9/01626 of the National Science Centre, Poland. This work is sponsored (in part) by the Chinese Academy of Sciences (CAS), through a grant to the CAS South America Center for Astronomy (CASSACA) in Santiago, Chile.

The ALMA projects were sponsored by the Universidad de Concepcion in Chile. This paper makes use of the following ALMA data: ADS/JAO.ALMA\#16.1.00794.S, ADS/JAO.ALMA\#2017.1.01689.S, ADS/JAO.ALMA\#2016.2.00033.S and ADS/JAO.ALMA\#2018.1.00047.S. ALMA is a partnership of ESO (representing its member states), NSF (USA) and NINS (Japan), together with NRC (Canada), MOST and ASIAA (Taiwan), and KASI (Republic of Korea), in cooperation with the Republic of Chile. The Joint ALMA Observatory is operated by ESO, AUI/NRAO and NAOJ. The National Radio Astronomy Observatory is a facility of the National
Science Foundation operated under cooperative agreement by Associated
Universities, Inc.

%

\vspace{5mm}
\facilities{ALMA; Crimea 1.25\,m}








\bibliography{ref}
\bibliographystyle{aasjournal.bst}



\begin{figure}[ht!]
\centering
\includegraphics[scale=0.48]{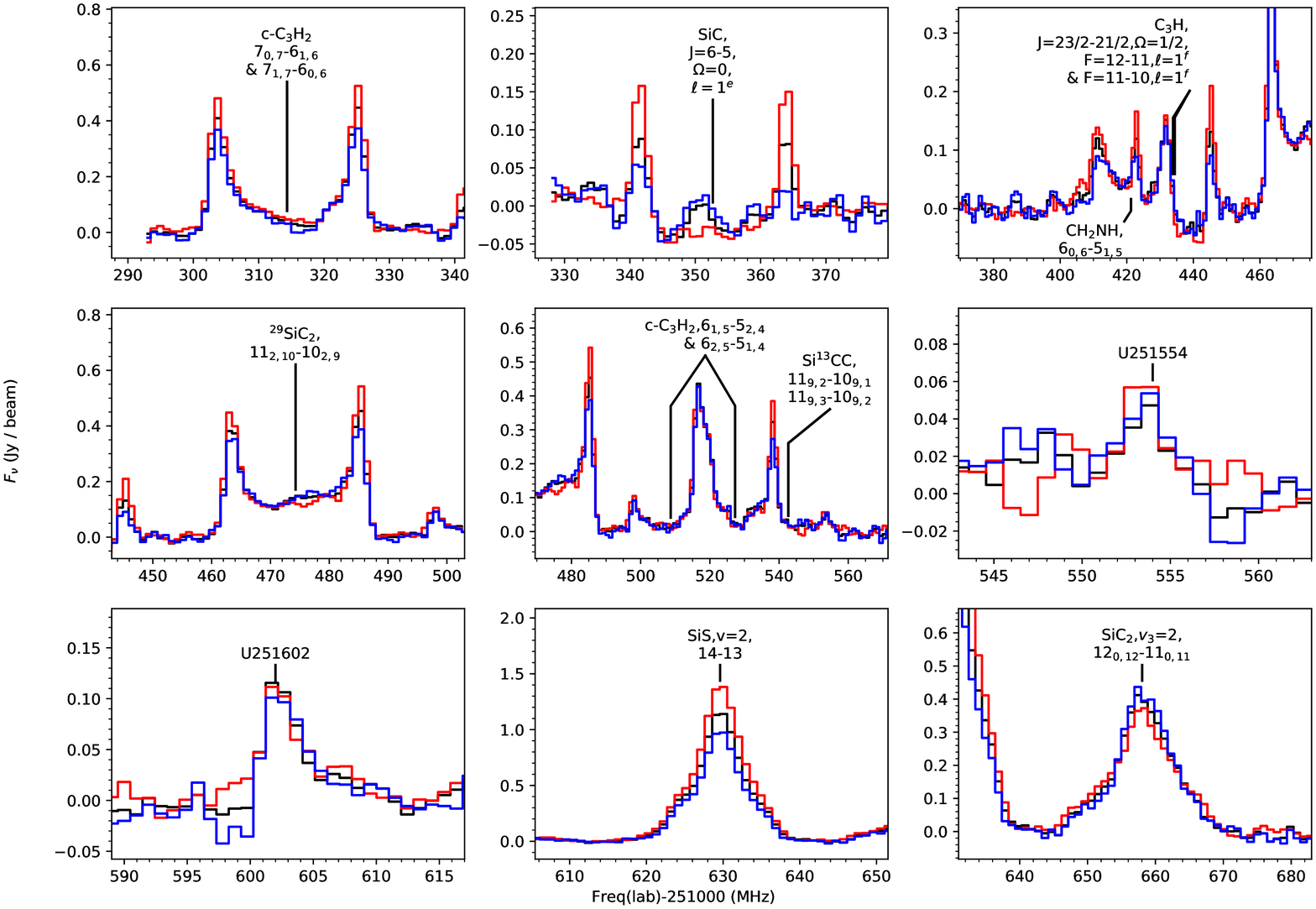}
\includegraphics[scale=0.48]{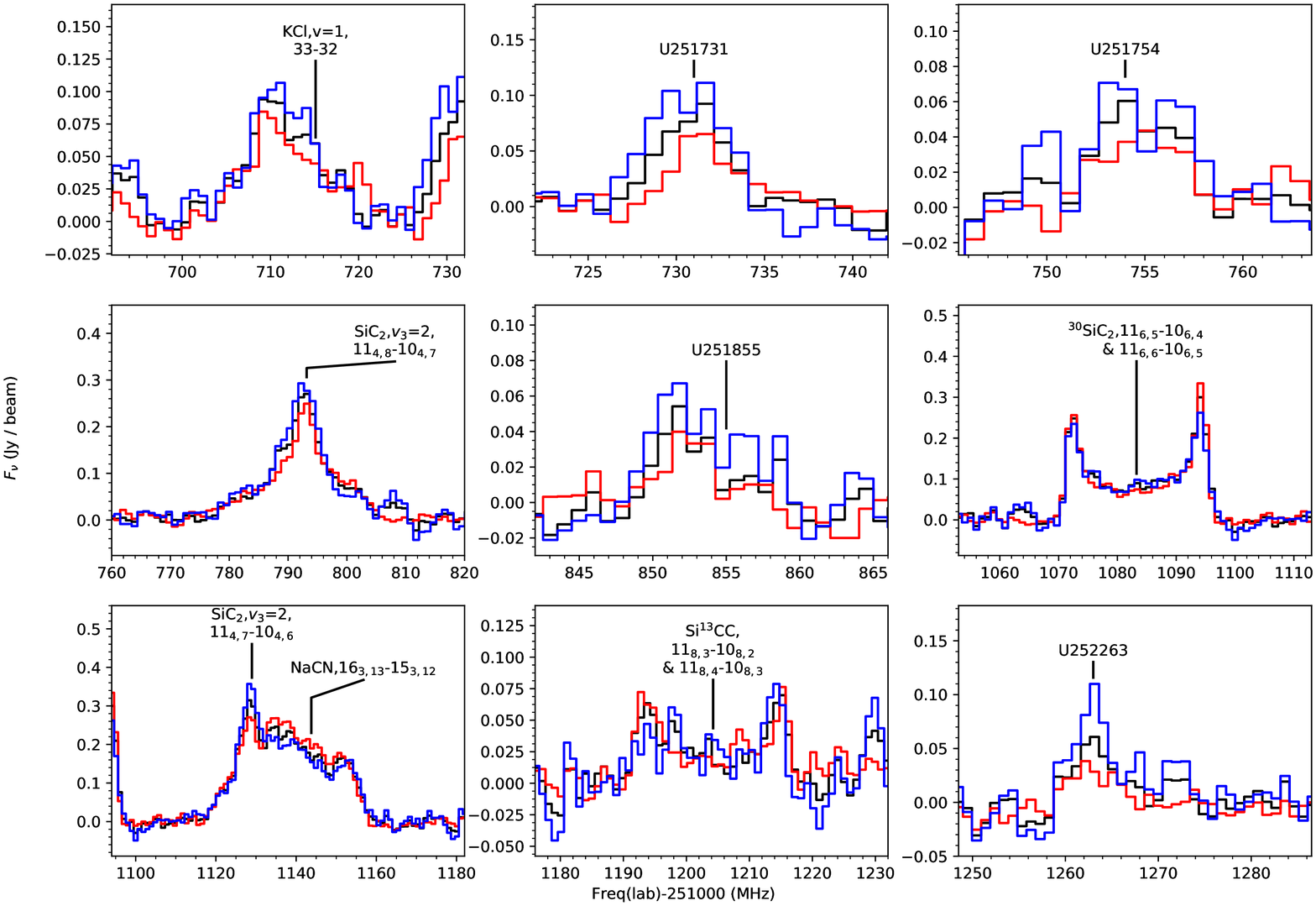}
\caption{Spectra of each group of line features that are averaged over the maximum epochs ({\em red}), minimum epochs ({\em blue}) and over a selected sample of epochs balanced between the maximum and minimum halves of the continuum light curve ({\em black}). The figure consists of two blocks, with each block composed of nine small panels. Each panel shows a group of neighboring line features that share the same spectral baseline. The lab frequency of the spectra in each figure block has been subtracted by a common reference frequency (shown in the title of abscissa) for brevity. \label{fig: compare-max-min1}}
\end{figure}
\begin{figure}[ht!]
\centering
\addtocounter{figure}{-1}
\includegraphics[scale=0.522]{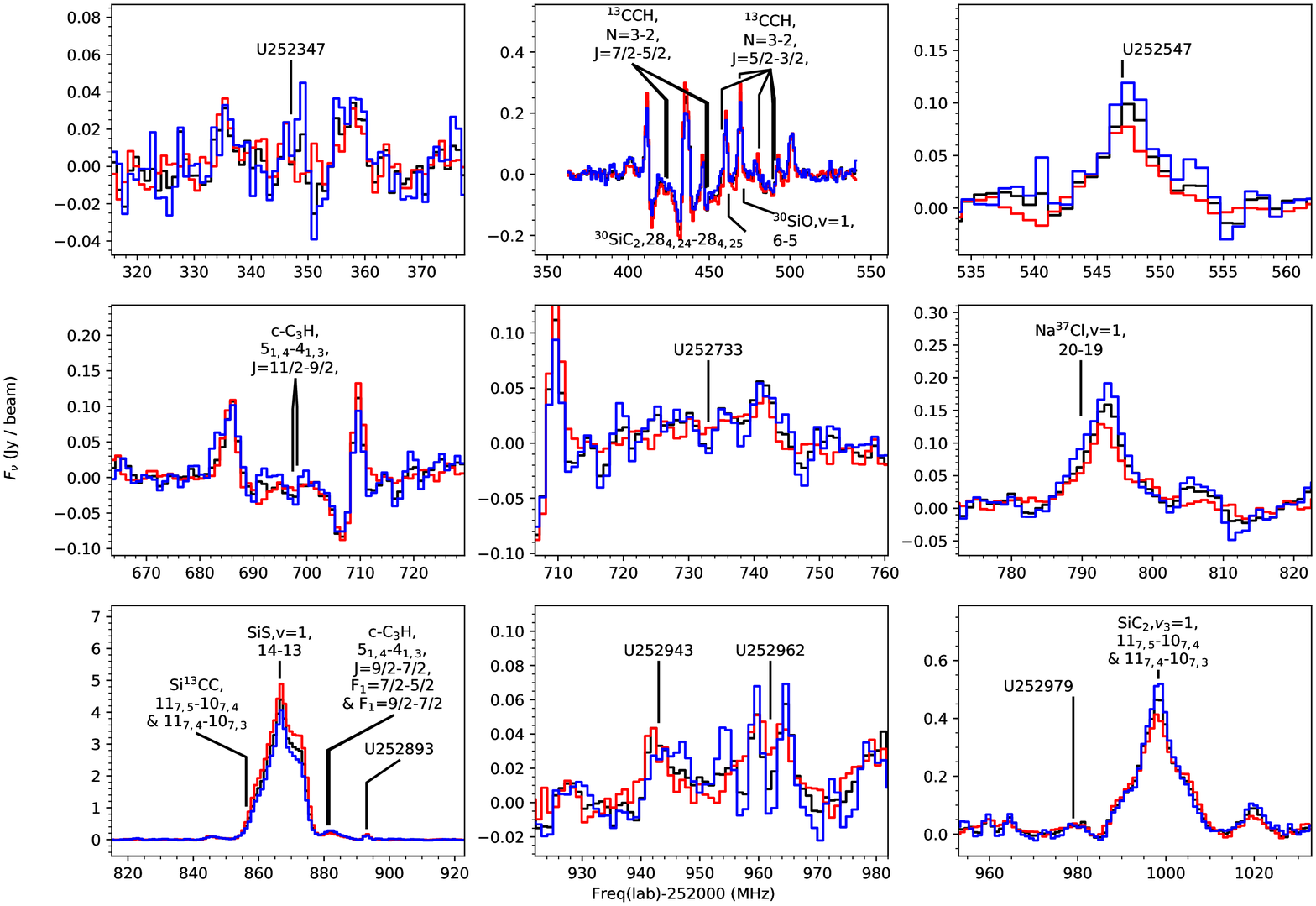}
\includegraphics[scale=0.522]{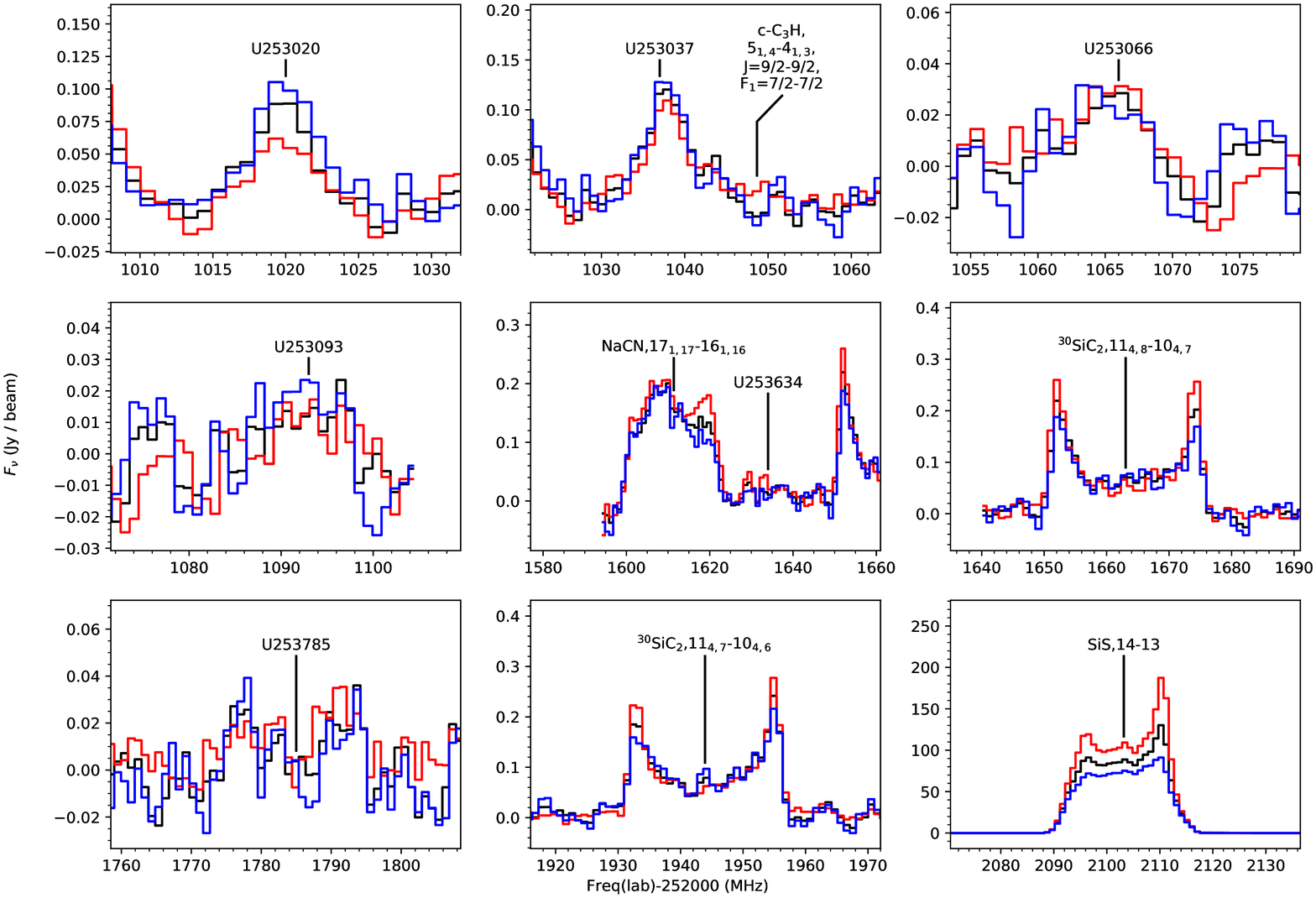}
\caption{(continued)}
\end{figure}
\begin{figure}[ht!]
\centering
\addtocounter{figure}{-1}
\includegraphics[scale=0.522]{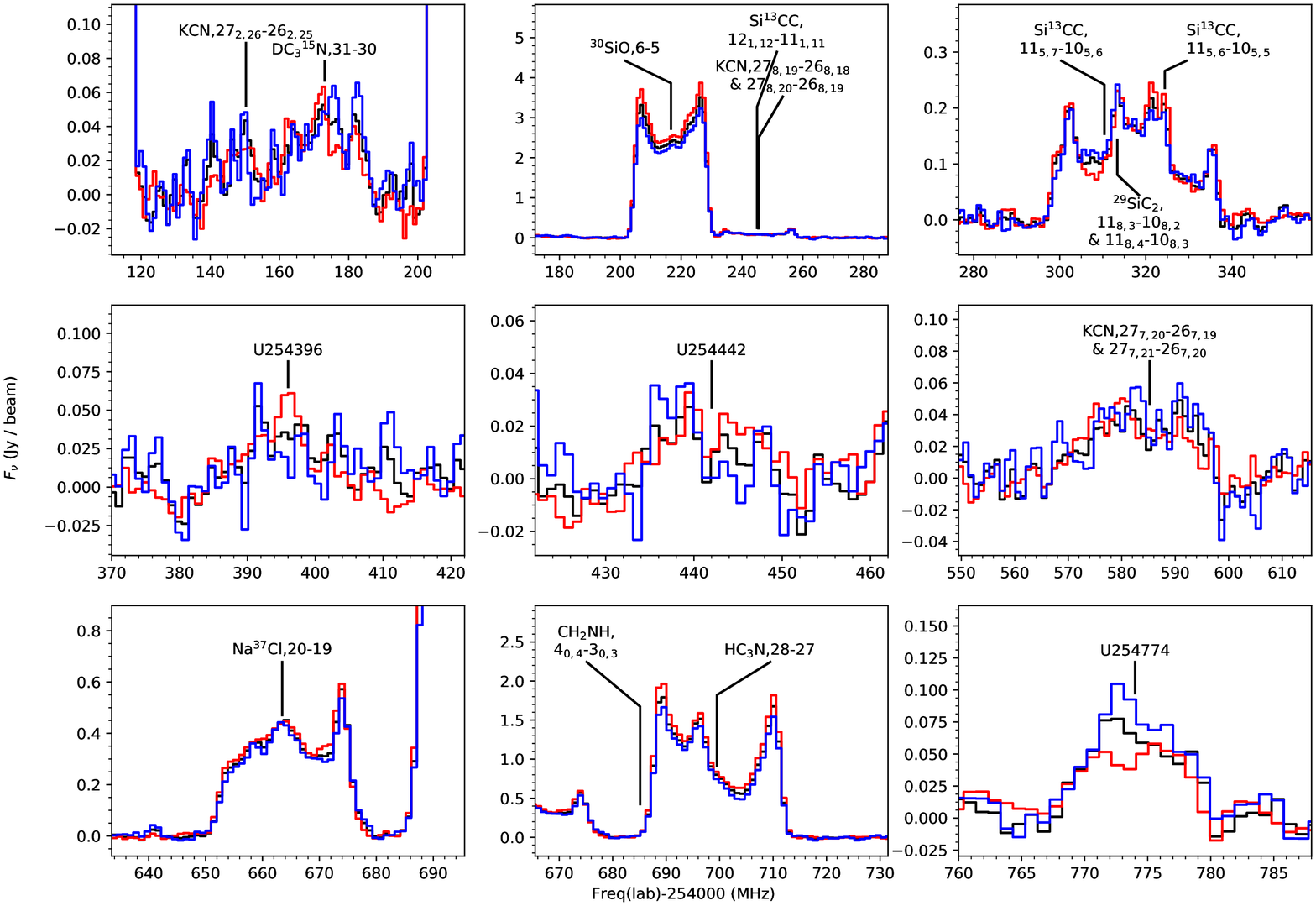}
\includegraphics[scale=0.522]{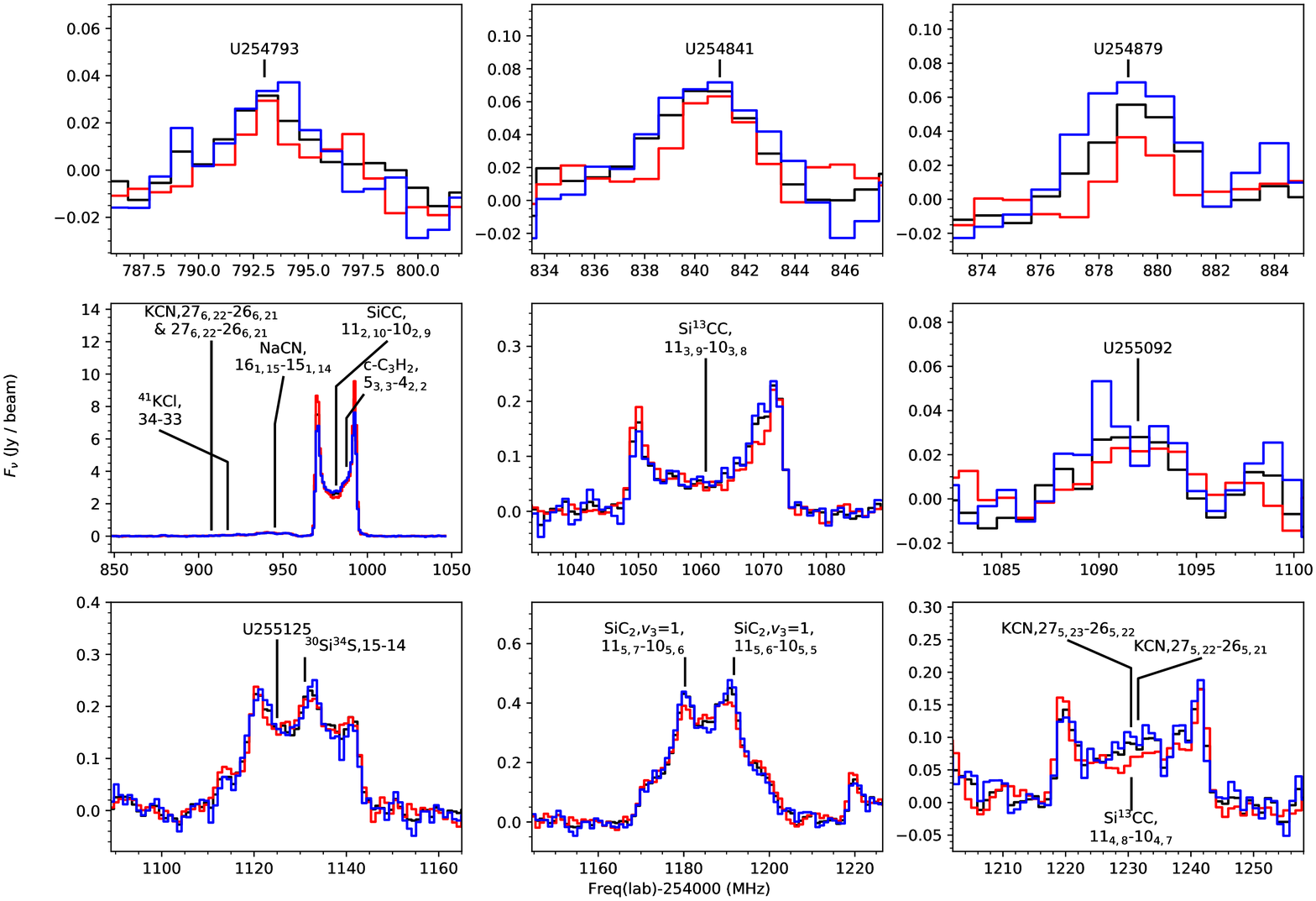}
\caption{(continued)}
\end{figure}
\begin{figure}[ht!]
\centering
\addtocounter{figure}{-1}
\includegraphics[scale=0.522]{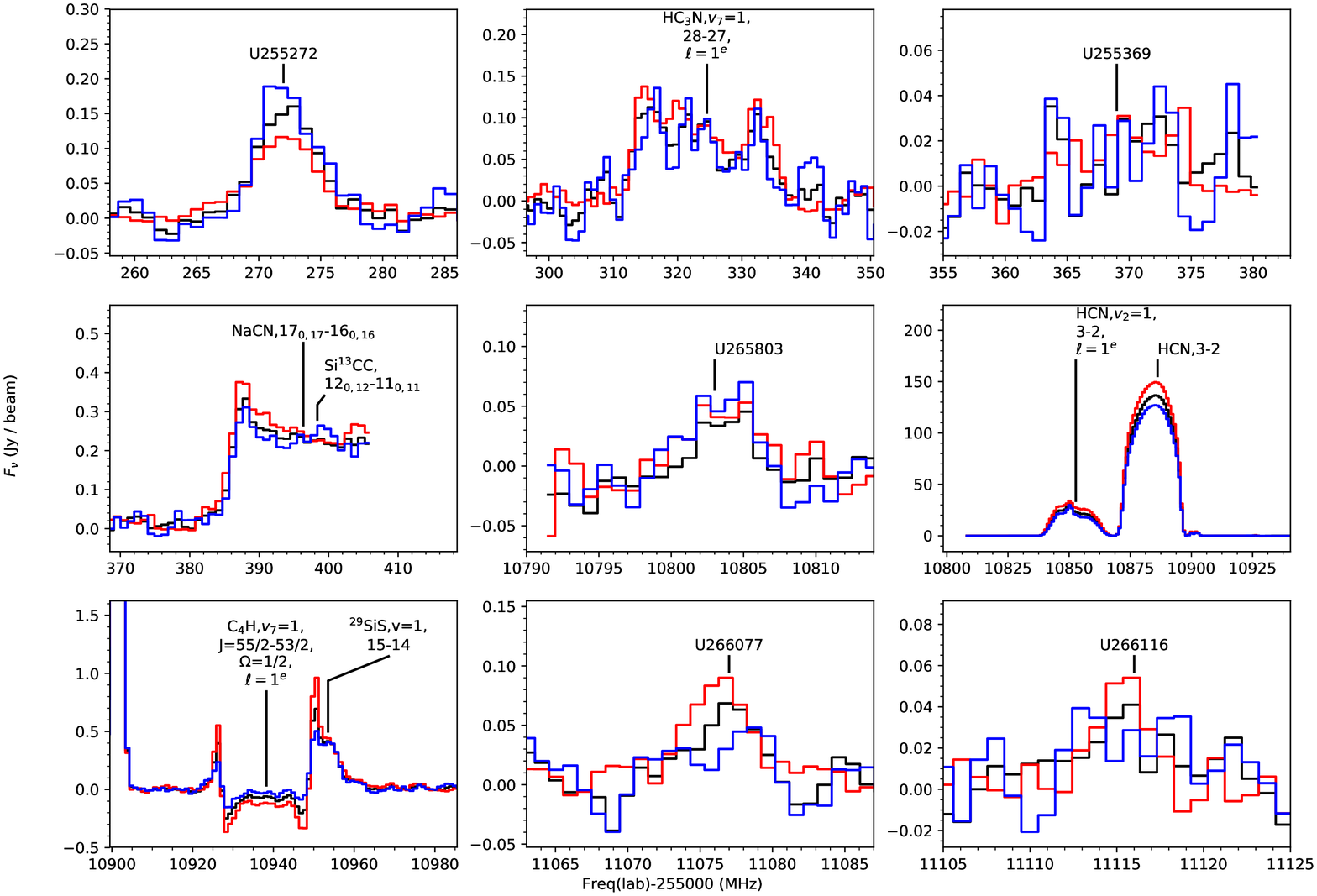}
\includegraphics[scale=0.522]{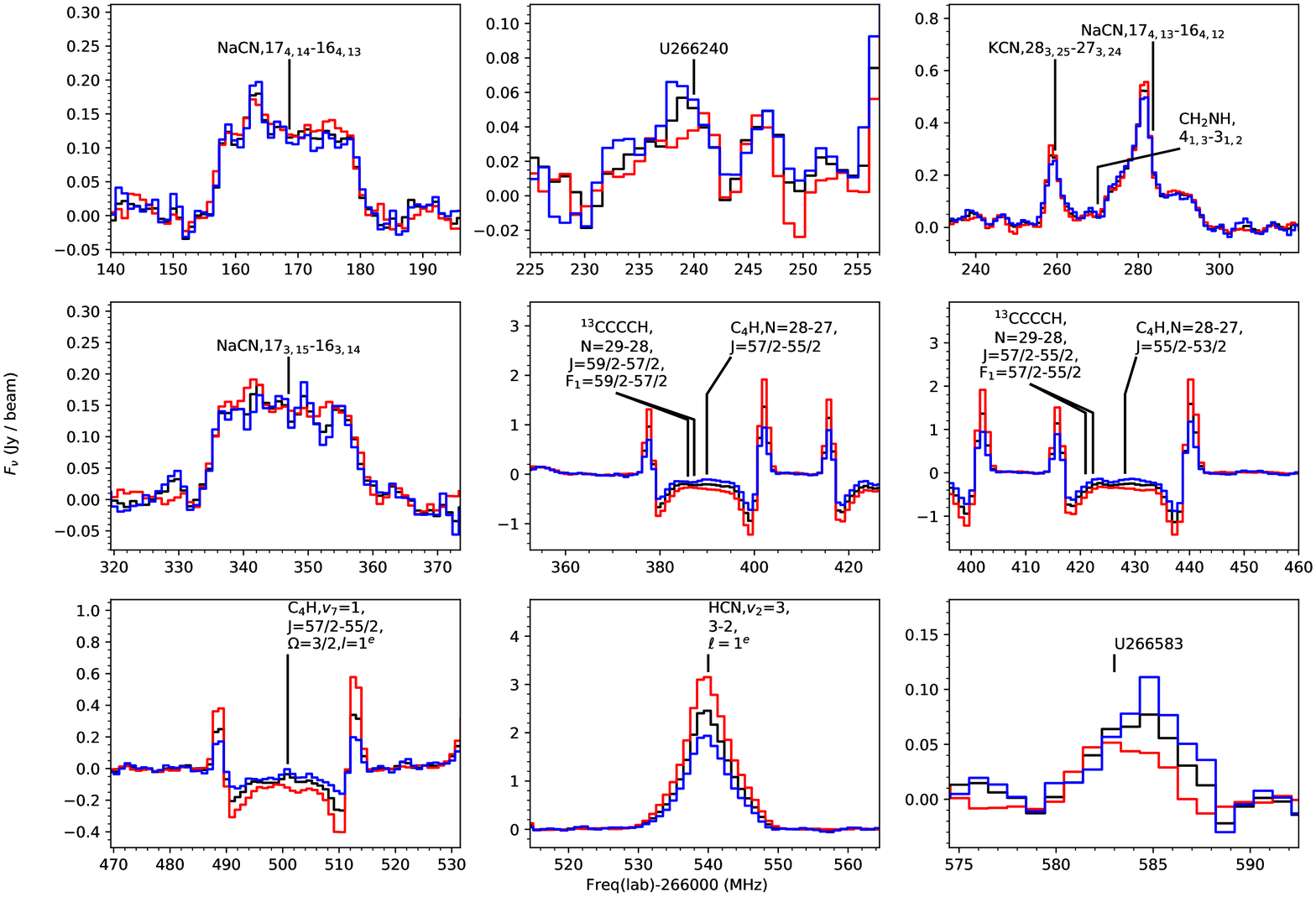}
\caption{(continued)}
\end{figure}
\begin{figure}[ht!]
\centering
\addtocounter{figure}{-1}
\includegraphics[scale=0.522]{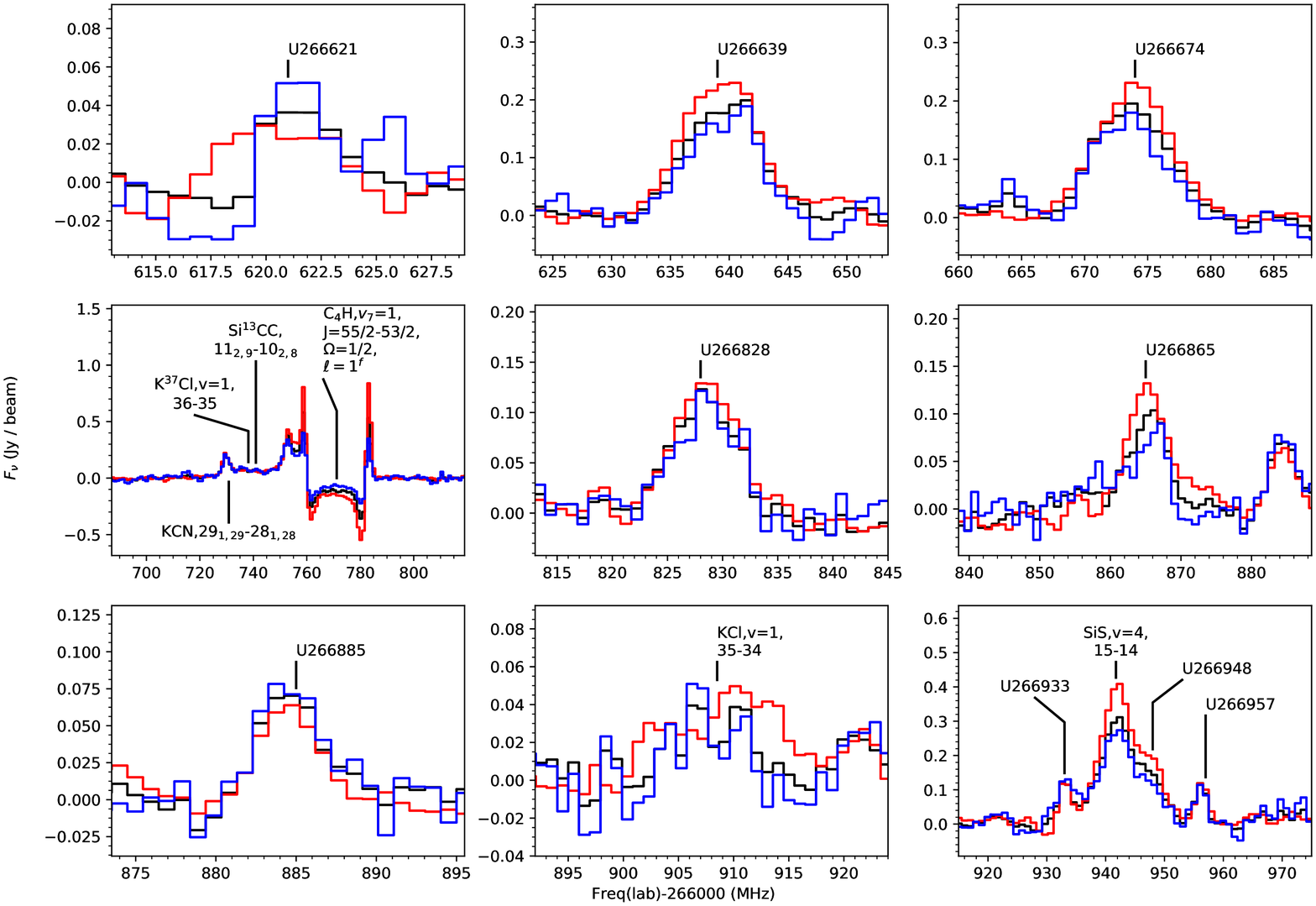}
\includegraphics[scale=0.522]{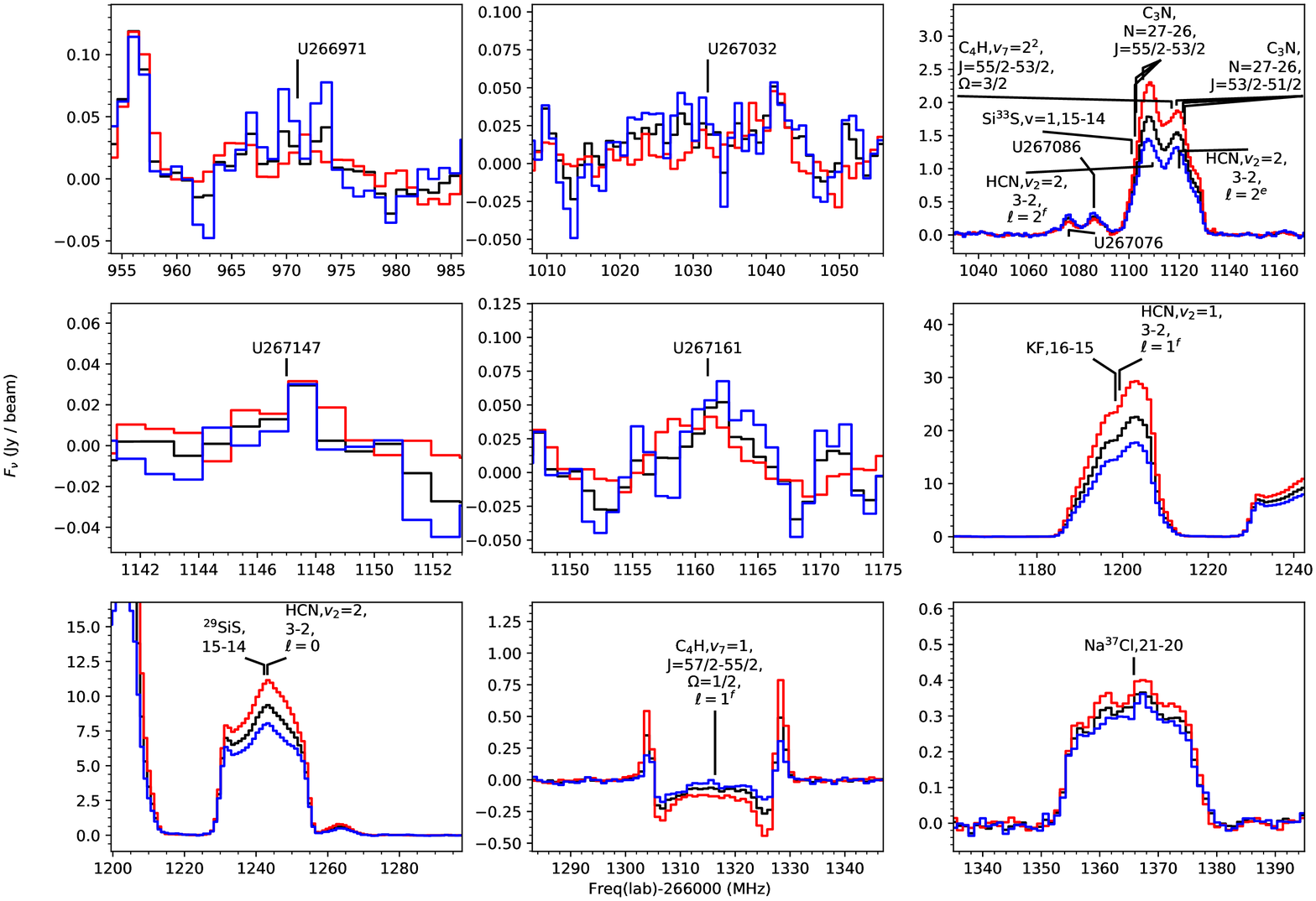}
\caption{(continued)}
\end{figure}
\begin{figure}[ht!]
\centering
\addtocounter{figure}{-1}
\includegraphics[scale=0.522]{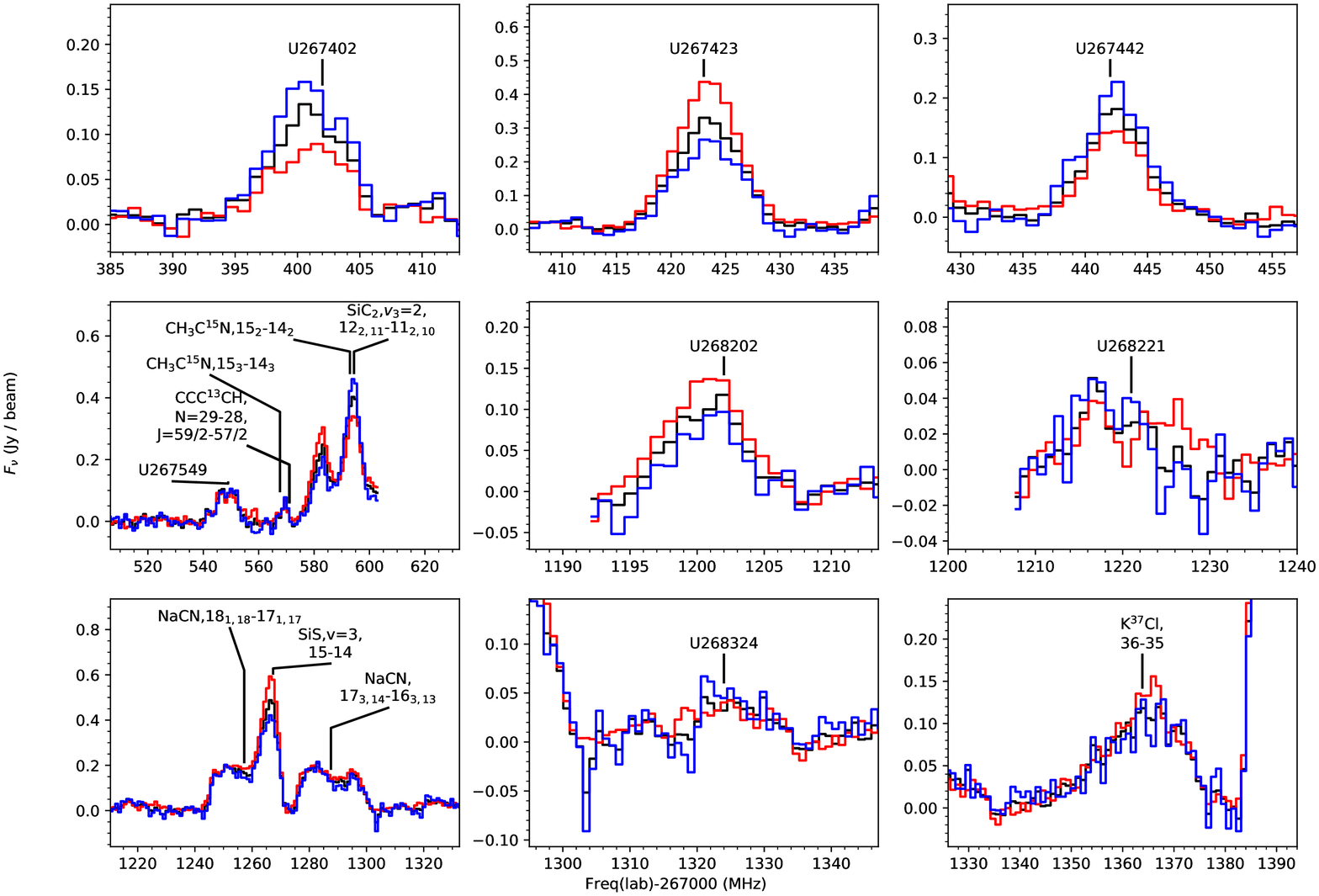}
\includegraphics[scale=0.522]{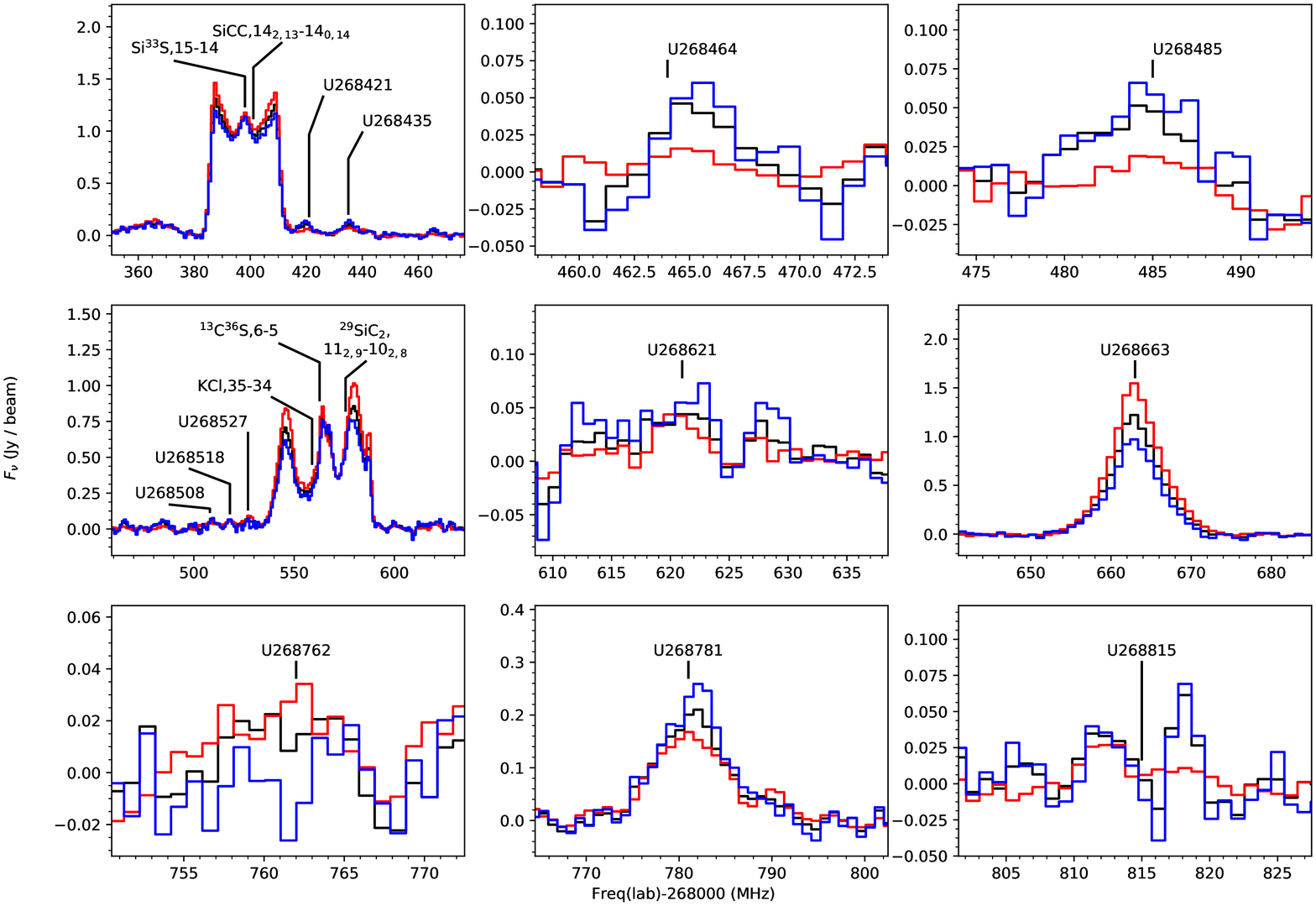}
\caption{(continued)}
\end{figure}
\begin{figure}[ht!]
\centering
\addtocounter{figure}{-1}
\includegraphics[scale=0.522]{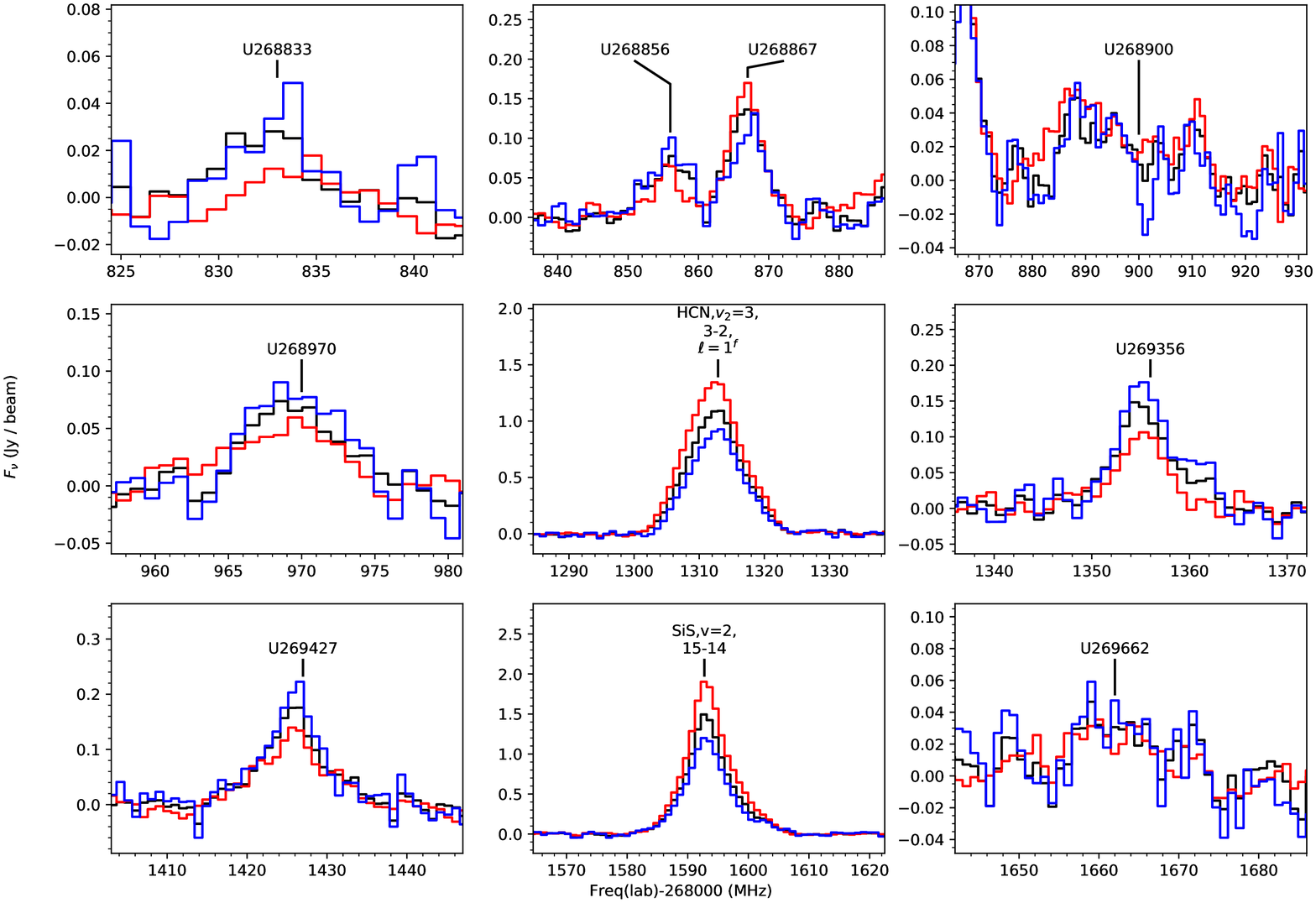}
\includegraphics[scale=0.522]{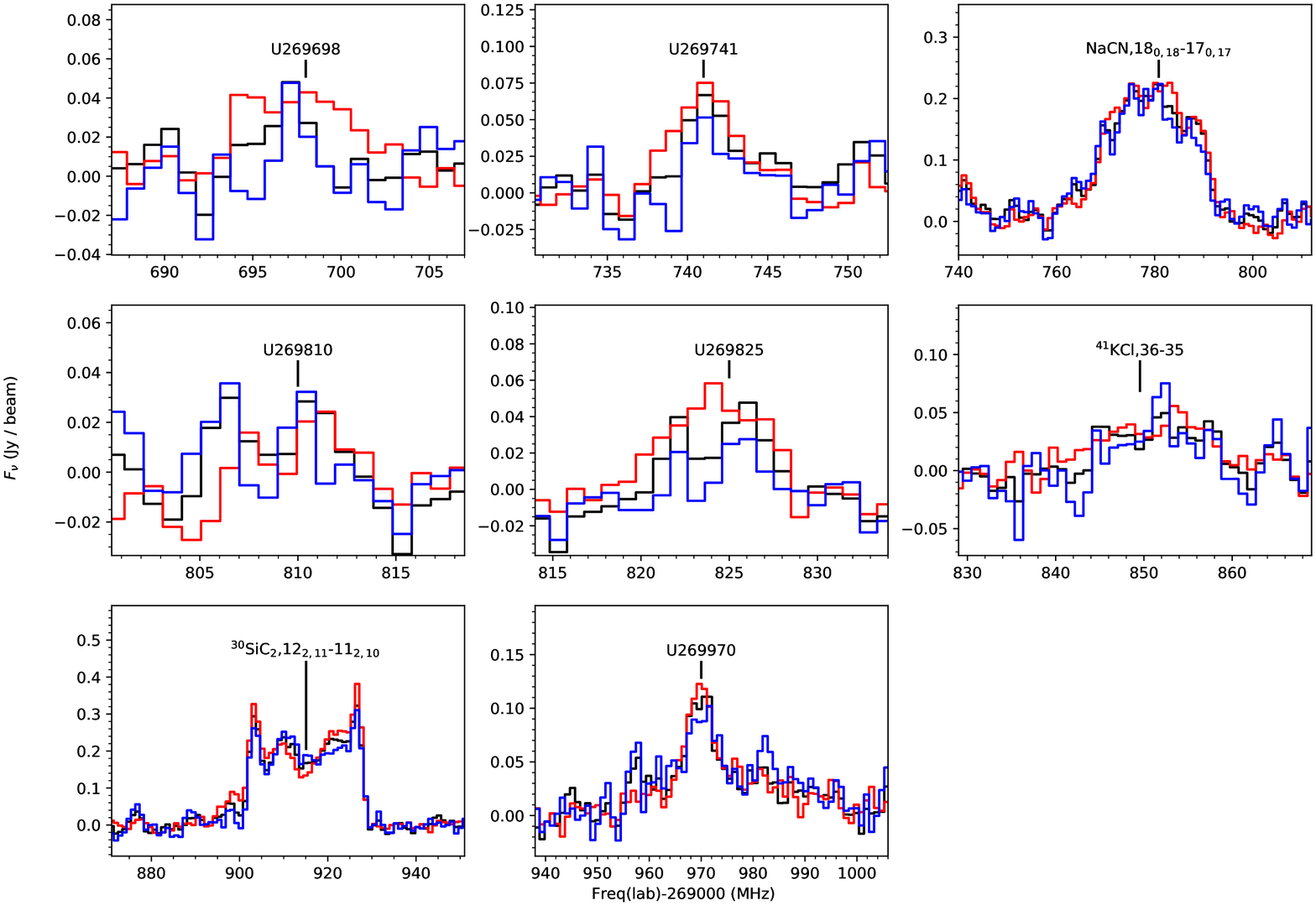}
\caption{(continued)}
\end{figure}

\end{document}